\documentclass[10pt]{article}
\usepackage{fullpage,epsfig,graphics,amsbsy,amssymb,cancel,bbm}
\usepackage{psfrag,hyperref,color,slashed}
\usepackage{graphicx}
\usepackage{amsfonts,mathdots,dsfont}
\usepackage{amssymb,amsmath,amscd,mathrsfs}
\usepackage{tikz}
\usetikzlibrary{arrows,chains,matrix,positioning,scopes,snakes}

\makeatletter
\tikzset{join/.code=\tikzset{after node path={%
\ifx\tikzchainprevious\pgfutil@empty\else(\tikzchainprevious)%
edge[every join]#1(\tikzchaincurrent)\fi}}}
\makeatother

\tikzset{>=stealth',every on chain/.append style={join},
         every join/.style={->}}
\tikzset{
    >=stealth',
    punkt/.style={
           rectangle,
           rounded corners,
           draw=black, very thick,
           text width=6.5em,
           minimum height=2em,
           text centered},
    pil/.style={
           ->,
           thick,
           shorten <=2pt,
           shorten >=2pt,}
}

\newcommand{\BB}{\mathbb}
\newcommand{\SF}{\mathsf}
\newcommand{\FR}{\mathfrak}

\def\L{{\cal L}}

\newcommand{\bea}{\begin{eqnarray}}
\newcommand{\eea}{\end{eqnarray}}
\newcommand{\nn}{\nonumber}
\newcommand{\Tr}{\textrm{Tr}}

\newcommand{\sbullet}{\textrm{\tiny{\textbullet}}}

\newcommand{\bra}{\langle}
\newcommand{\ket}{\rangle}

\newcommand{\im}{\textrm{Im}\,}
\newcommand{\re}{\textrm{Re}\,}
\newcommand{\To}{\Rightarrow}
\newcommand{\ind}{\textrm{ind}}
\newcommand{\sgn}{\textrm{sgn}}
\newcommand{\reeb}{\textrm{\scriptsize{$R$}}}
\newcommand{\sreeb}{\textrm{\tiny{$R$}}}
\newcommand{\gYM}{g_{\textrm{\tiny{$YM$}}}}

\def\ga{\alpha}
\def\gb{\beta}
\def\gc{\gamma}
\def\Gc{\Gamma}
\def\gd{\delta}
\def\ep{\epsilon}
\def\gt{\theta}

\def\gs{\sigma}

\def\gk{\kappa}
\def\gl{\lambda}
\def\Gl{\Lambda}
\def\Go{\Omega}
\def\go{\omega}

\DeclareMathAlphabet{\mathpzc}{OT1}{pzc}{m}{it}

\newcommand{\qed}{\nobreak \ifvmode \relax \else
      \ifdim\lastskip<1.5em \hskip-\lastskip
      \hskip1.5em plus0em minus0.5em \fi \nobreak
      \vrule height0.5em width0.5em depth0.00em\fi}

\def\spinc{spin${}^c~$}

\begin{document}
\thispagestyle{empty}
\begin{flushright} \small
UUITP-08/13
 \end{flushright}
\smallskip
\begin{center} \LARGE
{\bf 5D Super Yang-Mills on $Y^{p,q}$ Sasaki-Einstein manifolds}
 \\[12mm] \normalsize
{\bf  Jian Qiu$^a$ and Maxim Zabzine$^b$} \\[8mm]
 {\small\it
${}^a$Math\'ematiques, Universit\'e du Luxembourg,\\
 Campus Kirchberg, G 106,  L-1359 Luxembourg\\
      \vspace{.5cm}
${}^b$Department of Physics and Astronomy,
     Uppsala university,\\
     Box 516,
     SE-75120 Uppsala,
     Sweden\\
   }
\end{center}
\vspace{7mm}
\begin{abstract}
 \noindent
 On any simply connected Sasaki-Einstein five dimensional manifold one can construct a
  super Yang-Mills theory which
  preserves at least two supersymmetries.   We study  the special case of toric Sasaki-Einstein manifolds known as $Y^{p,q}$ manifolds. We use the localisation technique to compute the full perturbative part of the partition function.  The full equivariant result is expressed in terms of
   certain special function which  appears to be a curious generalisation of the triple sine function.
    As an application of our general result we study the large $N$ behaviour for the case of single
     hypermultiplet in adjoint representation and we derive the $N^3$-behaviour in this case.
  \end{abstract}

\eject
\normalsize
\tableofcontents
\section{Introduction}\label{sec_intro}

Recently the five dimensional Yang-Mills theory has attracted renewed attention and the interesting
 results were derived. These theories are
 interesting mainly due to their relations to, first of all, the magical $(2,0)$ 6D theory and
  secondly, the 5D $N=1$ SCFTs.
   Inspired by Pestun's work \cite{Pestun:2007rz} on localisation on $S^4$,
   many exact results have now been derived in diverse dimensions.
   For the case of 5D Yang-Mills theory on $S^5$, the partition function was derived and studied in a number of papers \cite{Kallen:2012va,Kim:2012av}, further extension to the squashed sphere was done in \cite{Imamura:2012xg,Lockhart:2012vp,Imamura:2012bm,Kim:2012qf}, and last but not least the case of $S^1\times S^4$ was studied in \cite{Kim:2012gu,Terashima:2012ra}, and other related background, see for example \cite{Fujitsuka:2012wg}.  These results were used
   in  providing checks  on the dualities AdS$_6$/CFT$_5$ \cite{Jafferis:2012iv, Assel:2012nf},
   AdS$_7$/CFT$_6$ \cite{Kim:2012av,Kim:2012qf,Minahan:2013jwa}  and as well as the AGT-inspired ideas \cite{Nieri:2013yra}.

In this work we go beyond the standard case of spheres and study the 5D susy gauge theory on a specific family of toric Sasaki-Einstein (SE) manifolds, the so called $Y^{p,q}$ manifolds (here $(p,q)$ are two coprime integers).  One major motivation is the intriguing results of Lockhart and Vafa \cite{Lockhart:2012vp}, which indicates how one may obtain the non-perturbative partition function from the purely perturvative part, as was inspired by the topological string considerations, thus we see it fit to extend our earlier computation on $S^5$ to a more intricate toric Sasaki-Einstein manifold, hoping that the non-trivial homology of $Y^{p,q}$ may provide more insight on the interpretation of the results.

Let us summarise briefly the results of the present paper. The six dimensional cone over $Y^{p,q}$
 can be obtained by the standard K\"ahler reduction of $\mathbb{C}^4$ under a $U(1)$ with charge $[p+q, p-q, -p, -p]$,
  where we use the standard notation for $U(1)$-actions on $\mathbb{C}^4$. Although
   only $U(1)^3$ acts on $Y^{p,q}$, it is convenient to discuss $U(1)^4$-actions on
   $\mathbb{C}^4$.  The full equivariant perturbative partition function for the 5D vector multiplet coupled to a hypermultiplet with mass $M$ in representation $\underline{R}$ has the following form
\bea
Z_{pert}=\int\limits_{\FR{t}}dx~\exp\big(-\frac{8\pi^3 r\varrho}{\gYM^2}\,\Tr[x^2]\big)
~\frac{{\det}'_{adj}S^{\Gl}(i x| \omega_1, \omega_2,  \omega_3, \omega_4)}{{\det}_{\underline{R}}S^{\Gl}(ix + iM + \frac{1}{2} (\omega_1 + \omega_2 + \omega_3 + \omega_4)| \omega_1, \omega_2, \omega_3, \omega_4)}~,\label{finale_intro}
\eea
where $\varrho =  \textrm{Vol}_{Y^{p,q}}/\textrm{Vol}_{S^5}$ (with $\textrm{Vol}_{Y^{p,q}}$ being the equivariant volume, see (\ref{volume_equiv}) for details) and   the function $S^{\Gl}$ is  defined as the zeta regularised infinite products
\bea
S^{\Gl}(x|  \omega_1, \omega_2, \omega_3, \omega_4)=\prod_{(i,j,k,l) \in \Gl^+}\Big(i \omega_1+j
\omega_2+k\omega_3+l\omega_4 +x\Big) \prod_{(i,j,k,l) \in \Gl^+_0}\Big(i\omega_1+j\omega_2+k\omega_3+l\omega_4 -x\Big) ~,\label{S_function-intro}\eea
where the lattices are defined as follows
\bea
&& \Gl^+=\big\{i,j,k,l\in\BB{Z}_{\geq0}\;|\;i(p+q)+j(p-q) -kp- lp=0\big\}~,\label{lattice-1}\\
&& \Gl_0^+=\big\{i,j,k,l\in\BB{Z}_{>0}\;|\;i(p+q)+j(p-q)-kp-lp=0\big\}~, \label{lattice-2}
\eea
 and $\omega_1, \omega_2, \omega_3, \omega_4$  are equivariant parameters corresponding to
  $U(1)^4$-action on $\mathbb{C}^4$.  The lattice conditions (\ref{lattice-1}) and  (\ref{lattice-2}) reduce the product to a three-dimensional lattice and $S^{\Gl}$ depends  effectively only on three parameters, i.e.
\bea
\big((p+q)\partial_{\go_1}+(p-q)\partial_{\go_2}-p\partial_{\go_3}-p\partial_{\go_4}\big)S^{\Gl}(x|\go_1,\go_2,\go_3,\go_4)=0~,\label{constraint}
\eea
reflecting the effective $U(1)^3$ action on $Y^{p,q}$-space. The function $S^{\Gl}(x|  \omega_1, \omega_2, \omega_3, \omega_4)$ resembles
   in many ways the triple sine functions and indicates how one may generalise the latter. The case of Sasaki-Einstein
    metric (or equivalently, the existence of two Killing spinors) on $Y^{p,q}$  space corresponds to the following specific choice of equivariant parameters
\bea
\omega_1=0~,~~~~\omega_2=\frac{1}{(p+q)\ell}~,~~~~\omega_3=\omega_4=\frac{3}{2}-\frac{1}{2(p+q)\ell}~.\nn
\eea
Following the analogy with $S^5$ we refer to this as unsquashed $Y^{p,q}$ space, whereas the case of arbitrary equivariant parameters will be called the squashed $Y^{p,q}$ space.
 In this paper we study the asymptotics of $S^{\Gl}$ for
 a general set of equivariant parameters (with some minor restriction (\ref{dual_cone_condition}) related to the moment cone and convergence).
  As a concrete application of our result we study the case of $SU(N)$ gauge theory
   with a single hypermultiplet in the adjoint representation. For the large $N$-limit, in the case of large
    't Hooft coupling the free energy behaves as follows
\bea
 F = - \frac{g_{YM}^2 N^3}{96 \pi r} \varrho \left ( \frac{9}{4} + M^2 \right )^2~,\label{free-energy-N}
\eea
 where $\varrho=\textrm{Vol}_{Y^{p,q}}/\textrm{Vol}_{S^5}$ and this for the case of unsquashed
  $Y^{p,q}$ space admitting Sasaki-Einstein metric. Thus we find the result
  which is identical to the calculation on $S^5$ up to the volume factor $\varrho$.

The paper is organised as follows: in section \ref{sec-YM} we review the construction of supersymmetric 5D Yang-Mills theory with matter on general Sasaki-Einstein manifolds.
 We also review some basic properties of the Saski-Einstein geometry.  Section \ref{sec_GoYpq} is a review of the definitions of $Y^{p,q}$, with some discussions on certain properties of $Y^{p,q}$ relevant for the subsequent index calculation.
Section \ref{sect-cohom} contains the localisation argument and discusses the relation between
 supersymmetry and cohomological complexes.
Section \ref{sec-computation} contains the technical calculation of one-loop determinants
 and the final result is given in terms of infinite products.
In section \ref{sec_asymptotic_analysis} we study the asymptotic behaviour of these infinite products.
 Using this asymptotic behaviour in section \ref{sec-matrix} we provide the application of our results
  for the case of single hypermultiplet in adjoint representation and we derive $N^3$-behaviour.
Section \ref{sec_discussion} summarises the results of the paper and indicates the possible generalisation for other toric Sasaki-Einstein manifolds. Many technical details and calculations
  are collected in Appendices.

\section{Super Yang-Mills Theory on Sasaki-Einstein 5-folds}
\label{sec-YM}

We start from the $N=1$ supersymmetric gauge theory on $S^5$ constructed in \cite{HosomichiSeongTerashima}, which has 8 super-charges, and place the theory on some other Sasaki-Einstein (SE) manifolds. Assuming that the SE manifold is simply connected, then one is guaranteed a pair of Killing spinors of type (1,1), and consequently a quarter supersymmetry (2 supercharges).  It turns out that without turning on more background fields from the  supergravity multiplet, only $S^5$ can have more super-charges, but on the other hand a quarter of the supersymmetry is sufficient for the purpose of localisation and we will not spend much effort in enlarging the supersymmetry, except some brief discussions in section \ref{sec_discussion}.

\subsection{5D Yang-Mills theory with matter on $S^5$}

The field content of the susy YM theory on $S^5$ consists of a vector-multiplet and a hyper-multiplet. The vector-multiplet contains
the gauge field $A_m$, a scalar $\gs$, an $SU(2)$-triplet of scalars $D_{IJ}$ and a symplectic Majarona gaugino $\gl_I$, with the following off-shell supersymmetry
  transformation (see (\ref{spinor_bi_linear}) for our notation of spinor bi-linears)
\bea
&&\gd A_m = i\xi_I\Gc_m\gl^I~,\nn\\
&&\gd\gs = i\xi_I\gl^I~,\nn\\
&& \gd\gl_I = -\frac12(\Gc^{mn}\xi_I)F_{mn}+(\Gc^m\xi_I)D_m\gs-\xi^JD_{JI}+\frac{2}{r} t_I^{~J}\xi_J\gs~, \label{susy_vect} \\
&&\gd D_{IJ} = -i\xi_I\Gc^mD_m\gl_J+[\gs,\xi_I\gl_J]+\frac{i}{r}t_I^{~K}\xi_K\gl_J+(I\leftrightarrow J)~,\nn
\eea
where $\xi_I$ is a spinor, satisfying the Killing equation
\bea
 D_m\xi_I=\frac{1}{r} t_I^{~J}\Gc_m\xi_J~,~~~t_I^{~J}=\frac{i}{2}(\sigma_3)_I^{~J}~,~~~(\xi_I\xi_J)=-\frac12\ep_{IJ}~,\label{killing_eqn}
 \eea
where $\sigma_3=\rm{diag}[1,-1]$. The quantity $t_I^{~J}$ is the vev of an $SU(2)_R$-triplet auxiliary field in the Weyl multiplet. We remark that in checking the closure property of the susy transformation, only the Killing spinor equation and the dimensionality of the space is used.

The Lagrangian density for the vector multiplet on $S^5$ is
\bea&& L_{vec}= \frac{1}{\gYM^2} \Tr\Big[\frac{1}{2}F_{mn} F^{mn} -D_m\gs  D^m\gs-\frac12 D_{IJ}D^{IJ}+\frac{2}{r} \gs t^{IJ}D_{IJ}- \frac{10}{{r}^2}
 t^{IJ}t_{IJ}\gs^2\nn\\
&&\hspace{2cm}+i\gl_I\Gc^mD_m\gl^I-\gl_I[\gs,\gl^I]-\frac{i}{r}t^{IJ}\gl_I\gl_J\Big]~,\label{action_vector}
\eea
where $F_{mn}$ is the field strength for $A_m$. The vector indices are raised and lowered with the metric, while the $SU(2)_R$-indices are raised using
$\epsilon^{IJ}$ (see Appendix  \ref{A-spinors}). The action is susy invariant, and also in checking this, only the Killing equation (\ref{killing_eqn}) is used.
 Therefore we can  take the supersymmetry transformations (\ref{susy_vect})  and  the action (\ref{action_vector}) from $S^5$ and use them on any simply connected SE manifold, but with of course only a fraction of the supersymmetry retained due to a smaller number of Killing spinor solutions.

The hyper-multiplet on $S^5$ consists of an $SU(2)_R$-doublet of complex scalars $q^A_I,~~I=1,2$ and an $SU(2)_R$-singlet fermion $\psi^A$, with the reality conditions ($A=1,2,\cdots,2N$)
\bea
 (q^A_I)^*=\Omega_{AB}\epsilon^{IJ}q^B_J~,~~(\psi^A)^*= \Omega_{AB}C\psi^B~,\label{reality-cond-spin}
 \eea
where $\Omega_{AB}$ is the invariant tensor of $Sp(N)$
\bea \Go=\left|\begin{array}{cc}
                                                          0 & \mathds{1}_N \\
                                                          -\mathds{1}_N & 0
                                                        \end{array}\right|,\nn\eea
and $C$ is the charge conjugation matrix.

The gauge group will be a subgroup of $Sp(N)$, in particular we consider the hyper-multiplet with the representation $\underline{N}\oplus\underline{\bar N}$ of $SU(N)$, which is embedded in $Sp(N)$ in the standard manner
\bea U\to \left|\begin{array}{cc}
            U & 0 \\
            0 & U^{-T}
          \end{array}\right|~,~~~U\in SU(N).\nn\eea
One can rewrite the scalar field $q$ into a more familiar form as
\bea q_1=\frac{1}{\sqrt2}\left|\begin{array}{c}
           \phi_+ \\
           \phi_-
         \end{array}\right|~,~~~q_2=\frac{1}{\sqrt2}\left|\begin{array}{c}
           -\phi_-^* \\
           \phi_+^*
         \end{array}\right|~,\label{rewrite_scalar}\eea
where $\phi_{\pm}$ transform in the $\underline{N}$ and $\underline{\bar{N}}$ of $SU(N)$ respectively. The fermion can be written in a similar
 manner
\bea \psi^A=\frac{1}2\left|\begin{array}{c}
                    \psi^\alpha \\
                    -C\psi^*_\beta
                  \end{array}\right|~,\label{rewrite_fermion}\eea
where $\psi^\alpha$ is now an unconstrained Dirac spinor transforming in $\underline N$ (here $\alpha$ is the index for the representation).
  Analogously we can discuss the adjoint representation of
 $SU(N)$ when two copies of the adjoint are embedded into that of $Sp(N)$.

Suppressing the gauge group index, the supersymmetry on-shell transformations are written as:
 \bea
 &&\delta q_I=-2i\xi_I\psi~,\nn\\
&&\delta\psi=\Gc^m\xi_I(D_mq^I)+i\sigma \xi_Iq^I-\frac{3}{r} t^{IJ}\xi_Iq_J~. \label{hyper-tran-noaux}
\eea
These transformations leave invariant the action with the following Lagrangian density
\bea
&&L_{hyp}=\epsilon^{IJ}\Omega_{AB}D_mq_I^A  D^m q_J^{B} -\epsilon^{IJ}q_I^A \gs_{AC} \gs^C_{~B}  q_J^B +
\frac{15}{2r^2} \epsilon^{IJ}\Omega_{AB}t^2  q_I^A  q_J^B
 \nn\\
&&\hspace{1cm}-2i \Omega_{AB}\psi^A\slashed{D}\psi^B-2\psi^A\gs_{AB} \psi^B -4\Go_{AB}\psi^A\gl_Iq^{IB}-iq_I^AD_{AB}^{IJ}q_J^B~,\label{action-matter-1}
\eea
where $t^2=t^{IJ}t_{IJ}=1/2$ and $\sigma_{AB} = \Omega_{AC} \sigma^C_{~B}$.
Here again, only the Killing spinor equation plus the Einstein relation $R_{mn}=4g_{mn}$ is used for the check.  A  mass term can be generated through the standard trick of coupling the hypermultiplet to an auxillary vector multiplet and giving an expectation value to the scalar in the multiplet, see
 \cite{HosomichiSeongTerashima}.  For the localisation we will need to have off-sheel realisation of supersymmetry transformations and for the hypermultiplet it will be resolved later on.

To summarise, with no modification, we have a supperymmetric Yang-Mills theory with matter on any SE manifold, with two supersymmetries. As we saw, the existence of supersymmetry depends on the existence of Killing spinors.  In the next two subsections, we quickly go over some necessary facts about  SE manifolds and the explicit construction of Killing spinors. This material is quite well-known by now and thus our review is mainly to set the notations. The reader may consult the short but nice review \cite{BoyerGalickishort}.

\subsection{Sasaki-Einstein Manifolds}
A {\it contact metric structure} on $M^5$ consists of a 1-form $\gk$ such that $\gk(d\gk)^2\neq0$ and a complex structure $J$ on the sub-bundle $\ker\gk$, which we call the \emph{horizontal plane}
\bea
J\in\textrm{Aut}(\ker\gk)~,~~~~~J^2=-\mathds{1}~,\nn\eea
and that $J$ is compatible with $d\gk$ in the sense that
$1/2d\gk J$ is a metric for the horizontal plane $\ker\gk$.
One can choose a unique vector field $\reeb$ such that
\bea
\iota_{\sreeb}\gk=1~,~~~~~\iota_{\sreeb}d\gk=0~,\nn\eea
and we extend $J$ to act also on $\reeb$ as zero $J\reeb=0$, leading to
\bea
J^2=-\mathds{1}+{\reeb}\otimes\gk~.\nn
\eea
The metric of the tangent bundle is the direct sum of the one on $\ker\gk$ and the one along ${\reeb}$
\bea
g=\frac1{2}d\gk J+\gk\otimes\gk~.\nn
\eea
As a consequence
\bea
&&g(JX,JY)=g(X,Y)-\gk(X)\gk(Y)~,\nn\\
&&d\gk=-2gJ~,\label{kappa_J}\\
&&{\reeb}=g^{-1}\gk~.\nn\eea

If ${\reeb}$ is a Killing vector field, then $(\gk,{\reeb},J)$ gives a \emph{K-contact structure}, the Killing condition is equivalent to
\bea\nabla_X{\reeb}=JX.\label{nabla_v}\eea

A \emph{Sasaki-manifold} is a K-contact manifold such that its metric cone $M\times (0,\infty)$ with metric and symplectic form
\bea G=r^2g+dr^2,~~~\Go=d(r^2\gk)~,~~~~~{\cal J}=2\Go^{-1}G~.\nn
\eea
is K\"ahler. The complex structure is written explicitly as
\bea
{\cal J}=J+r^{-1}\reeb\otimes dr-r\partial_r\otimes \gk~,\nn
\eea
it is easy to check ${\cal J}^2=-\mathds{1}_6$. The vector field
\bea \ep=r\frac{\partial}{\partial r}\label{homothetic}\eea
is called the \emph{homothetic} vector field, and it is clear that
\bea
 {\cal J}(\ep)=\reeb~.\nn
\eea

The K\"ahler condition is equivalent to the covariant constancy of ${\cal J}$ with respect to the Levi-Civita connection
\footnote{It can be shown that the closedness of the wouldbe K\"ahler form $G{\cal J}$ plus the vanishing of the Nijenhuis tensor is equivalent to the covariant constancy of ${\cal J}$, see lemma 4.15  in \cite{McDuffSalamon}}.
Thus a K-contact manifold is Sasaki iff $J$ satisfies the
integrability condition
\bea
&&\bra Z,(\nabla_XJ)Y\ket=-\gk(Z)\bra X,Y\ket+\bra Z,X\ket\gk(Y)\label{integrability}~,
\eea
where $\bra-,-\ket$ is the inner product using the metric. From now on we will use the \emph{same letter $J$ for the complex structure as well as the 2-form $gJ$}.

If the cone metric is in addition Ricci-flat i.e. the cone is actually Calabi-Yau, then $M$ is said to be Sasaki-Einstein (SE). The Ricci flatness is equivalent to
\bea
R_{mn}=4g_{mn}~.\label{Einstein}
\eea
Using the Reeb, one can define the \emph{horizontal} forms
\bea \go\in\Go^{\sbullet}_H(M)~,~~~~\textrm{if}~~~~\iota_{\sreeb}\go=0~.\nn\eea
Let us fix the volume form of $M^5$ as\footnote{Due to a historical accident, the choice of volume form in \cite{Kallen:2012va} is opposite to the current one, see also footnote  \ref{footnote_minus_reeb}. The reader should bear this in mind when comparing results between the two papers, especially some anti-self-dualities there will become self-dualities here.}
\bea
 \textrm{vol}=\frac12\kappa\wedge J\wedge J~,\label{volume_form}
 \eea
and one can define a duality for the horizontal 2-forms as
\bea
\omega\to *_{\sreeb}\omega=\iota_{\sreeb}*\omega,~~~\omega\in\Go_H^2(M^5)~,\label{horizontal_dual}\eea
Next one can prove that for an SE manifold, the Weyl tensor is horizontal and anti-selfdual
\bea \iota_{\sreeb}W_{XY}=0~,~~~*_{\sreeb}W_{XY}=-W_{XY}~,\label{Weyl}\eea
where
\bea
W_{XY}=X^mY^n\big(R_{mnpq}-g_{p[m}g_{n]q}\big)~,~~~~X,Y\in TM~.\label{def_weyl}
\eea
The proof makes use of (\ref{integrability}) and the details can be found in chapter 5 of \cite{DavidEBlair}.

\subsection{Sasaki Einstein 5-folds and Killing Spinors}\label{sec_killing_spinor}

Using the horizontal complex structure one can define the so called canonical \spinc-structure.
Let
\bea
W_{can}=\small{\textrm{$\bigoplus$}}\,\Omega_H^{0,\sbullet}(M)~,\label{can_spin_bundle}
\eea
where $\Omega_H^{0,\sbullet}$ consists of horizontal forms anti-holomorphic with respect to $J$.
One then has a representation of the Clifford algebra: let $\psi$ be any section of $W_{can}$ and $\chi$ a 1-form, define the Clifford action
\bea
\chi\cdotp\psi=\Bigg\{
                             \begin{array}{cc}
                               \sqrt2 \chi\wedge\psi & \chi\in\Omega_H^{0,1}(M) \\
                               \sqrt2 \iota_{g^{-1}\chi}\psi & \chi\in\Omega_H^{1,0}(M) \\
                               (-1)^{\deg+1}\psi & \chi=\kappa \\
                             \end{array}.\label{can_spin_rep}\eea
In this way, one has a \spinc-structure whose characteristic line bundle (see chapter 5 in \cite{Salamon}) is the anti-canonical line bundle associated with the complex structure $J$.
If $M$ is actually SE, then condition (\ref{Einstein}), (\ref{kappa_J}) together with the condition $H_1(M,\BB{Z})_{tor}=0$ would imply that $M$ is spin (theorem 7.5.27  in \cite{BoyerGalicki}).

For a 5D SE spin manifold one can show that there exists a pair of killing spinors satisfying
\bea
D_m\xi^1=-\frac{i}{2}\Gc_m\xi^1~,~~~~D_m\xi^2=+\frac{i}{2}\Gc_m\xi^2~,\nn
\eea
we will review the construction from \cite{FriedrichKath}. Consider the following dimension 1 sub-bundle $W_0$ within the spin bundle $W$
\bea
\psi\in W_0\subset W~,~~~{\reeb}\psi=-\psi~,~~~\frac12(1+i\mu J)X^{\perp}\cdotp\psi=0~,~~~\forall X\in\Gc(TM)~,\label{sub_bundle}
\eea
where $\mu=\pm1$ and to keep the formulae neat, \emph{we have omitted $\Gc$ whenever the Clifford multiplication is obvious}. In view of the construction (\ref{can_spin_rep}), the condition above says that $\psi$ is in $\Go_H^{0,0}(M)$ or  $\Go_H^{0,2}(M)$ depending on $\mu$.
One can rewrite the second condition in (\ref{sub_bundle}) into
\bea
A(X)\psi=\Big(\mu JX-\frac {i}{2}{\reeb}X-\frac i2X\Big)\psi=0~.\nn
\eea
One then defines a connection for the subbundle $W_0$
\bea
\tilde D_X=D_X+\frac{i\mu}{2}X~,\nn
\eea
and checks that this is indeed a connection, i.e.
\bea
(1)~ [\tilde D_X,{\reeb}]\psi=0~,~~~~(2)~[\tilde D_X,A(Y)]\psi=0~,~~~~\psi\in W_0~.\nn
\eea
The crucial step now is to show that the \emph{curvature of $\tilde D$ is zero when restricted to $W_0$}. First an explicit calculation shows that the curvature is given by the Weyl-tensor
\bea
[\tilde D_X,\tilde D_Y]-\tilde D_{[X,Y]}=\frac14R_{XYpq}\Gc^{pq}-\frac14[X,Y]=\frac14X^mY^n(R_{mnpq}-g_{p[m}g_{n]q})\Gc^{pq}=\frac14W_{XYpq}\Gc^{pq}~.\nn
\eea
which was shown to be horizontal and anti-self-dual. In fact, for any such 2-form $Z_{pq}$ one has $\slashed{Z}\psi=0$ for $\psi\in W_0$
\bea \slashed{Z}\psi=Z_{pq}\Gc^{pq}\psi=-\frac{\sqrt{g}}{2} Z_{rs}\ep^{rs}_{~~pqt}\reeb^t\Gc^{pq}\psi
=Z^{rs}\Gc_{rst}\reeb^t\psi=Z^{rs}(\Gc_{rs}\Gc_t-2\Gc_rg_{st})\reeb^t\psi=-\slashed{Z}\psi~,\label{aselfdualpsi}\eea
where we used $\iota_{\reeb}Z=0$ and (\ref{duality_curv}). To summarise the foregoing arguments, we have reached the conclusion:

\noindent
\emph{If a Sasaki-Einstein manifold $M^5$ is simply connected, then the solution to the Killing equation
\bea D_X\psi=-\frac{i\mu}{2}X\,\psi~,~~~~\mu=\pm1\nn
\eea
exists and is unique up to a constant scale factor.} The simply connectedness is needed to ensure we have no non-trivial flat bundle. Moreover, the solution satisfy
\bea
&& {\reeb}\psi=-\psi~,~~~~\big(\mu JX-\frac{i}{2}(1+{\reeb})X\big)\psi=0~,\label{subbundle}\\
&&\slashed{J}\psi=-4i\mu\psi~,\label{spin_form_degree}\eea
where the second line is a simple consequence of the first.

\section{Geometry of $Y^{p,q}$ Manifolds}
\label{sec_GoYpq}

In this section we briefly review some relevant facts about the family of toric SE manifolds known as $Y^{p,q}$ manifold.  For further details the reader may consult \cite{Boyer_talk,Martelli:2004wu,BoyerPati}.

\subsection{$Y^{p,q}$ from Reduction}
%
%

Take $\BB{C}^4$ with coordinates $[z_1,z_2,z_3,z_4]$ and the standard complex structure. Let
 us introduce the vector fields $e_i$ and one forms $\eta_i$
\bea
e_i=i(z^i\partial_{z^i}-c.c)~,~~~~\eta_i=\frac{i}{2}(z^id\bar z^i-c.c)~,~~~~\eta=\sum_i \eta_i~,\label{contact_1_form}
\eea
 where $e_i$ generates the phase rotation of $z_i$ and $\eta_i$ is its dual 1-form.
  It is clear that $d\eta$ gives the standard K\"ahler form on $\BB{C}^4$.

Consider now a $U(1)_T$ acting on $\BB{C}^4$ with charge $[p+q,p-q,-p,-p]$, the vector field generated by the action is\footnote{Here we follow the convention from \cite{Boyer_talk}, but the explicit metric given in
\cite{Gauntlett:2004yd} corresponds to the charge vector $[p,p,-(p-q),-(p+q)]$.}
\bea
T=(p+q)e_1+(p-q)e_2-pe_3-pe_4~.\label{T}
\eea
We perform a K\"ahler reduction with respect to $T$.
The moment map for $T$ is
\bea \mu_T=(p+q)|z_1|^2+(p-q)|z_2|^2-p|z_3|^2-p|z_4|^2~,\nn\eea
and the reduction
\bea C(Y^{p,q})=\mu_T^{-1}(0)/U(1)_T\nn\eea
is again K\"ahler and has a cone structure due to the special form of $\mu_T$. This will be the K\"ahler cone over a base which is by definition a Sasaki-manifold. One can get the base of the cone by imposing
$\sum\limits_i a_i|z_i|^2=1$, with $a_i\gneq0$. One possible choice is that of a squashed $S^7$
\bea
(p+q)|z_1|^2+(p-q)|z_2|^2+p|z_3|^2+p|z_4|^2=1~,\label{squashed_S7}
\eea
which together with $\mu_T=0$ leads to $S^3\times S^3$
\bea
\mu^{-1}_T(0)\big|_{S^7}=\{(z_1,z_2,z_3,z_4)\;|\;(p+q)|z_1|^2+(p-q)|z_2|^2=1/2,~~p|z_3|^2+p|z_4|^2=1/2\}\sim S^3\times S^3~.\nn\eea
Of course, there is nothing special about the choice (\ref{squashed_S7}), if one chooses instead $\sum_i|z_i|^2=1$, one still gets $S^3\times S^3$.
We remark that the $T$ action is free on $S^3\times S^3$ if $\gcd(p,q)=1$, since $z_{1,2}$ cannot be zero together nor can $z_{3,4}$. So $Y^{p,q}$ can also be presented as a quotient of $S^3\times S^3$ by $T$.

With the choice (\ref{squashed_S7}), one has the Reeb vector field
\bea
\reeb_1=(p+q)e_1+(p-q)e_2+pe_3+pe_4~,\label{Reeb}
\eea
it is easy to check that on $Y^{p,q}$, we have $\iota_{\reeb_1}\eta=1$, and $\iota_{\reeb_1}d\eta=0$.
Moreover $\eta$ will descend to $Y^{p,q}$ and be the contact 1-form $\gk$ there. This Reeb will not admit an SE metric, but it does help us to get a better handle on the geometry of our SE manifold.

\smallskip
\noindent $\sbullet$ The orbifold Base of $Y^{p,q}$.

In contrast to $T$, the Reeb vector field (\ref{Reeb}) does not in general induce a free action, even though it is nowhere zero. Indeed, if one takes the quotient of $Y^{p,q}$ with respect to
  $\reeb_1$, one gets the weighted projective space
\bea \BB{C}P(p+q,p-q)\times \BB{C}P(1,1)~.\nn\eea
In fact, this is easier to see if one takes the double quotient of $S^3\times S^3$ with respect to
both $T$ and $\reeb_1$.
The spaces $\BB{C}P(r,s)$ are orbifolds except when $r=s$. This special case happens when $p=1$ and $q=0$, which leads to the SE manifold called $T^{1,1}$, whose cone is the well-known conifold
\bea |z_1|^2+|z_2|^2-|z_3|^2-|z_4|^2=0~.\label{conifold}\eea
In fact $T^{1,1}$ is a regular SE manifold, since the Reeb orbit is closed and thus $T^{1,1}$ is a $U(1)$-fibration over $S^2\times S^2$ of degree 1 and 1.

For an orbifold, we can still apply some version of the Riemann-Roch theorem, and our computation of the super-determinant in subsection \ref{sec_method_I} is similar to this in spirit.
In the next section, we will instead look at the possibility of presenting $Y^{p,q}$ as a $U(1)$-fibration. Our result will concur with the geometric interpretation of the explicit metric found in \cite{Gauntlett:2004yd}.

\subsection{Looking for Free $U(1)$ in $Y^{p,q}$}\label{sec_LfFUiY}
Next we try to find free $U(1)$-actions in this geometry. Let $\ga$ be the $U(1)$ with charge $[a,c,b,d]$
\bea \begin{array}{c|cccc}
          & z_1 & z_2 & z_3 & z_4 \\
         \hline
         T & p+q & p-q & -p & -p \\
         \ga & a & c & b & d \\
         \reeb_1 & p+q & p-q & p & p
       \end{array}\nn\eea
The result is that if $\gcd(p,q)=1$, then a free $U(1)$ has charge vector
\bea [a,c,b,d]=[a,-a-2b,b,b]~,~~~~\textrm{where}~~a,b,c,d\in\BB{Z}~,~~~(a+b)p+bq=1~.\label{solu_free_action}\eea

For the proof, note first that for $\ga$ to be free, it can be nowhere parallel with $T$, otherwise it will have a zero on the quotient. Even when this is satisfied, at special points there might still be discrete stability groups.
For example, when $z_2=z_4=0$, if (where $k,l\in\BB{Z}$)
\bea &&\frac{\gt}{2\pi}a=\frac{\phi}{2\pi}(p+q)+k~,\nn\\
&&\frac{\gt}{2\pi}b=\frac{\phi}{2\pi}(-p)+l~,\nn\eea
then a rotation $e^{i\gt \ga}$ can be undone by $e^{-i\phi T}$. So if $\gt/(2\pi)$ and $\phi/(2\pi)$
have rational (but non-integer) solutions, then there is a non-trivial stability group. The solution to these equations is given by
\bea \frac{\gt}{2\pi}=\frac{pk+(p+q)l}{ap+b(p+q)}~,\nn\\
\frac{\phi}{2\pi}=\frac{-bk+al}{ap+b(p+q)}~.\nn\eea
To exclude any non-integer solution, we need to have $\gt/(2\pi)$ to be an integer for \emph{any} $k$ and $l$, but as $p,\;q$ are coprime, one can find $k,\;l$ s.t. $kp+l(p+q)=1$, thus the denominator $ap+b(p+q)$ must be $\pm1$. And similar reasonings lead to three more equations
\bea \bigg\{
       \begin{array}{c}
         (a+b)p+bq=\pm1 \\
         (a+d)p+dq=\pm1 \\
       \end{array}~,\nn\\
     \bigg\{
       \begin{array}{c}
         (c+b)p-bq=\pm1 \\
         (c+d)p-dq=\pm1 \\
       \end{array}~.\nn\eea
Let us first assume that the rhs of the first pair is $+1$, which implies $b=d$, and that the rhs of the second pair must also have the same sign. Take this first to be $+1$, then we have
\bea
(a+b)p+bq=1~,~~~~(c+b)p-bq=1~,\nn
\eea
which would have a solution only when $p=1,2$, so we do not consider this possibility. Now take the $-1$ option
\bea
 (a+b)p+bq=1~,~~~(c+b)p-bq=-1~,\nn
 \eea
which admits a family of solutions
\bea [a,c,b,d]=[a,-a-2b,b,b]~,~~~~(a+b)p+bq=1~.\nn
\eea
One possibility left is when the right hand side of the first pair have $+1,-1$
\bea \bigg\{
       \begin{array}{c}
         (a+b)p+bq=+1 \\
         (a+d)p+dq=-1 \\
       \end{array}~,\nn\eea
then subtracting the two equations one get $(b-d)(p+q)=2$, which again admits no solutions since $p>q>0$. Thus we have only the set of solutions (\ref{solu_free_action}) that works for generic $p,~q$.

\subsection{The base of the $U(1)$ fibration}\label{sec_U(1)_base}
We investigate the base of the $U(1)$ fibration by a taking further quotient of $Y^{p,q}$ by $\ga$.
To this end, one may consider modding out $S^3\times S^3$ by
any $SL(2,\BB{Z})$ combinations of the two $U(1)$'s $T$ and $\ga$. We choose
\bea
 -aT+(p+q)\ga=[0,-2,1,1]~,~~~-bT-p\ga=[-1,1,0,0]~,~~~~\det\Big(
                                                         \begin{array}{cc}
                                                           -a & p+q \\
                                                           -b & -p \\
                                                         \end{array}
                                                       \Big)=1~.\label{convenient_SL2Z}
                                                       \eea
This shows that the quotient consists of a twisted product $S^2\rtimes S^2$. Indeed, from the first equation, one sees that the first $S^2$ is fibred over the second one with degree $-2$, as such we will call the two $S^2$'s the \emph{fibre} and \emph{base} $S^2$ respectively. From the second equation, the complex structure of the fibre $S^2$ is quite unconventional, if we choose it to agrees with the standard one at $z_2=0$ then at $z_1=0$, it will be opposite, see section 5.3 in \cite{Martelli:2004wu} for more discussion on this fact.

We can cover the base of the $U(1)$-bundle with four patches, let $U_{00}$ denote the patch with $z_2\neq0$, $z_4\neq0$, and $U_{10}$ be the patch $z_1\neq0$, $z_4\neq0$, etc.
For later computation of the index, we pick on patch $U_{00}$ two vector fields $e_3=[0,0,1,0]$ and $e_1=[1,0,0,0]$, and we tabulate the expression of these vector fields in other patches (change of coordinates)
\bea \begin{array}{l|c|c}
       \hline
         & e_3 & e_1 \\
       \hline
       U_{00} & [0,0,1,0] & [1,0,0,0] \\
       U_{01} & [2,0,0,-1]+(q-p)\ga-(a+2b)T & [1,0,0,0] \\
       U_{10} & [0,0,1,0] & [0,1,0,0]+p\ga+bT \\
       U_{11} & [0,2,0,-1]-aT+(p+q)\ga & [0,1,0,0]+bT+p\ga \\
       \hline
     \end{array}\label{equi_action}\eea
where $T$ of course descends to zero on $Y^{p,q}$. In the appendix, we will see that this change of coordinates can be precisely reproduced by analyzing the explicit SE metric.


\subsection{The Reeb that Admits a Sasaki-Einstein Metric}\label{sec_TRtAaSEM}
The above Reeb vector field will lead only to a Sasaki structure on the manifold, but not an Einstein metric. In a beautiful paper \cite{Martelli:2005tp}, it is shown how to find the Reeb vector that admits an SE metric, which for the computation of partition function is sufficient.

The SE manifold $Y^{p,q}$ correspond to what is known as being of the \emph{Reeb type}, meaning that it is a torus fibration over a base and that the Reeb vector field is generated by the torus action. Let us again start from $\BB{C}^4$ with the standard K\"ahler structure given by $d\eta$. Performing the symplectic reduction as before
\bea \BB{C}^4//T=\mu_T^{-1}(0)/T~,~~~~\mu_T=(p+q)|z_1|^2+(p-q)|z_2|^2-p(|z_3|^2+|z_4|^2)~,\nn\eea
one obtains a cone over a base $Y^{p,q}$. This cone inherits a K\"ahler structure from $\BB{C}^4$, which will be \emph{held fixed} in the deformation to come later. One can choose the following effectively acting 3-tori
\bea
T_1:~[0,0,1,0]~,~~~T_2=[-1,0,1,0]~,~~~T_3=\ga~,\label{effective_U(1)}\eea
there is quite some freedom in the choice of the tori, here we are following \cite{Martelli:2004wu}. It is easy to write down the moment map
\bea
 \mu_1=|z_3|^2~,~~~~\mu_2=|z_3|^3-|z_1|^2~,~~~~p\mu_3=|z_1|^2-|z_2|^2~,\nn
 \eea
where for the third relation, we used the second equation of (\ref{convenient_SL2Z}). From $|z_i|^2\geq0$, the range of $\mu_i$ is a polytope cone, described as
\bea
{\cal C}=\{[\mu_1,\mu_2,\mu_3]\,|\,\vec \mu\cdotp \vec v_a\geq0\}~,\label{poly_cone}
\eea
where $\vec v_a,~a=1\cdots 4$ are the inward pointing normal of the facets of the cone
\bea
\vec v_1=[1,0,0]~,~~~~\vec v_2=[1,-2,-p+q]~,~~~~\vec v_3=[1,-1,-p]~,~~~~\vec v_4=[1,-1,0]~.\nn
\eea
Then all possible Reeb vectors must live in the interior of the cone ${\cal C}^*$ generated by $\vec v_a$, i.e.
\bea
\reeb=b_1T_1+b_2T_2+b_3T_3~,~~~~\vec b=\sum \gl_a\vec v_a~,~~\gl_a>0~.\nn\eea
This is easy to see, let $\mu_{\reeb}$ be the moment map corresponding to $\reeb$, then $\mu_{\reeb}=$const will intersect the cone ${\cal C}$ at a polygon precisely when $\vec b\in({\cal C}^*)^{\circ}$. This polygon will serve as the base of the torus fibration mentioned earlier.

In deforming the Reeb vector, one holds the K\"ahler form fixed and deform the complex structure and hence the K\"ahler metric of the cone.
One can write down the Einstein Hilbert action of the metric
\bea S[\reeb]=\int_{Y^{p,q}}d\mu~(R-12)~,\nn\eea
where $R$ is the Ricci scalar and the stationary point of the action gives the Einstein metric $R_{mn}=4g_{mn}$.
The insight from \cite{Martelli:2005tp} is that the action only depends on the Reeb vector and the action has a unique stationary point within the interior of the dual cone ${\cal C}^*$. Quite remarkably, the value of the Hilbert-Einstein action can be written as some simple elementary geometrical quantities of the moment cone ${\cal C}$, which can in turn be related to the volume of the $Y^{p,q}$ space
\bea
S[\reeb]=8(b_1-2)\textrm{Vol}\,Y^{p,q}[\reeb]~,\label{Einstein_Hilbert}\eea
where $\textrm{Vol}\,Y^{p,q}[\reeb]$ is a function of the Reeb. With our choice of basis of $U(1)$'s in (\ref{effective_U(1)}), the parameters $b_i$ are related to the general equivariant parameters as
\bea
b_1=\omega_1 + \omega_2 + \omega_3 + \omega_4~,~~~~
b_2=-\omega_1-\omega_2 -2\omega_4~,~~~~b_3=-p\omega_2+(q-p)\omega_4~,\nn\eea
and the volume is computed to be
\bea
\varrho[\reeb]=\frac{\textrm{Vol}\,Y^{p,q}[\reeb]}{\textrm{Vol}\,S^5}=\frac { \left( {p}^{2}\omega_1+{p}^{2}\omega_2+{p}^{2}\omega_3+{p}^{2}\omega_4-pq\omega_1+pq\omega_2-{q}^{2}\omega_3-{q}^{2}\omega_4 \right) p} {\left( p\omega_1+p\omega_4+q \omega_4 \right)  \left( p\omega_2+p\omega_4- q\omega_4 \right)
 \left( p\omega_2+p\omega_3-q \omega_3 \right)  \left( p\omega_1+p\omega_3+ q \omega_3 \right)}~.\label{volume_equiv}\eea
 The condition that $\vec b$ be within the dual cone translates to some conditions on $\go_i$
\bea p(\go_2+\go_4)>q\go_4,~~p(\go_2+\go_3)>q\go_3,~~p(\go_3+\go_1)+q\go_3>0,~~p(\go_4+\go_1)+q\go_4>0,\label{dual_cone_condition}\eea
as a by product, one sees that any free $U(1)$ found in subsection \ref{sec_LfFUiY} will necessarily be disqualified as a Reeb.

To continue, one can then find the unique Reeb vector field admitting the SE metric by finding the stationary point of  (\ref{Einstein_Hilbert}), the result is
\bea \reeb=\big(\frac{3}{2}-\frac{1}{2(p+q)\ell}\big)(e_3+e_4)+\frac{1}{(p+q)\ell}e_2~,\label{decomp_of true_reeb}\eea
where
\bea \ell=\frac{q}{3q^2-2p^2+p\sqrt{4p^2-3q^2}}=\frac{2p^2-3q^2+p\sqrt{4p^2-3q^2}}{9q(p^2-q^2)}~.\label{def_ell}\eea
This is exactly the one found using the explicit metric from \cite{Gauntlett:2004yd}. One can plug back the value of $\reeb$ and get the volume of $Y^{p,q}$ with the SE metric
\bea \varrho= \frac{\textrm{Vol}_{Y^{p,q}}}{\textrm{Vol}_{S^5}}=\frac{q^2[2p+(4p^2-3q^2)^{1/2}]}{3p^2[3q^2-2p^2+p(4p^2-3q^2)^{1/2}]}=\frac{1}{27p^2(q^2-p^2)}\big(-8p^3+9pq^2-(4p^2-3q^2)^{3/2}\big)~.\label{volume}\eea
As for $Y^{p,q}$ with a general $\reeb$, we will call it the \emph{squashed} $Y^{p,q}$, whose volume is given by (\ref{volume_equiv}).

Notice that $\ell$ here appears in the explicit metric as the period of the fibre coordinate of the $U(1)$-fibre over the base $S^2\rtimes S^2$, see appendix \ref{sec_tEM}.
Moreover, since $\ell$ is generically irrational, the resulting Reeb vector field is of an \emph{irregular} type, i.e. its orbit is not closed.
But for the $T^{1,1}$ case $\ell^{-1}$ goes to zero when one sets $p=1$ and $q=0$, and the Reeb vecotr $\reeb=3/2[0,0,1,1]$ is freely acting and has closed orbit.

\section{The Cohomological Complex and Localisation}\label{sect-cohom}
The key step of the localisation procedure is to make a change of variable in the fields, so that the fields would behave as coordinates and their conjugate momenta (both even and odd) on some space. In this process, a combination of the susy transformation will behave like the equivariant differential, and thus one has the standard localisation in equivariant cohomology.

The process of change of variable is given in \cite{Kallen:2012va} and here we will just sketch the steps.

\subsection{Vector-Multiplet}
We first define some geometrical quantities using the Killing spinor. First, define the vector field
\bea
 \reeb^p=-\xi_I\Gc^p\xi^I~.\nn\eea
The Killing spinor equation implies that $\reeb$ is a Killing vector field, in fact, it is the Reeb vector field of our SE geometry \footnote{The vector field $-\sreeb$ takes the place of $v$ used in \cite{Kallen:2012va}. \label{footnote_minus_reeb}}, from the way our Killing spinor is constructed in subsection\ref{sec_killing_spinor}.
Furthermore, the spinor bi-linear
\bea t^{IJ}\xi_I\Gc_{mn}\xi_J=-\frac{1}{2}J_{mn}\nn\eea
is the horizontal K\"ahler 2-form \footnote{Recall that we use the same letter for the horizontal complex structure and the K\"ahler 2-form.}. It satisfies
\bea \iota_{\sreeb}J=0~,~~~~~*_{\sreeb}J=J~,\nn\eea
where $*_{\sreeb}$ is defined in (\ref{horizontal_dual}).

The gaugino $\gl_I$ in (\ref{susy_vect}) can be converted into a 1-form and a 2-form:
\bea \Psi_m=\xi_I\Gc_m\gl^I~,~~~~~\chi_{mn}=\xi_I\Gc_{mn}\gl^I+\reeb_{[m}   \xi_I\Gc_{n]}\gl^I~,\label{field_redef}\eea
where $\reeb_m=g_{mn}\reeb^n$ is the contact 1-form $\gk$. The 1-form $\Psi$ is an unrestricted 1-form, the 2-form $\chi$ satisfies the same conditions as $J$:
\bea \iota_{\sreeb}\chi=0~,~~~~\iota_{\sreeb} * \chi=\chi~.\label{horizontal_self_dual}\eea
The formula (\ref{field_redef}) can be inverted to write $\gl_I$ as
\bea\gl_I=-\frac12\xi^J(\xi_J\Gc^{mn}\xi_I)\chi_{mn}+(\Gc^m\xi_I)\Psi_m~,\label{redefine_spinor_fine}
\eea
The fermion $\gl_I$ has 8 real components which is the same as the
5 components from $\Psi$ plus 3 more from $\chi$.

In the new variables the susy transformations (\ref{susy_vect}) can be rewritten as
\bea
&&\begin{array}{ll}
  \delta A_m = i\Psi_m~, & \gd \Psi_m = -\reeb^n F_{nm}+D_m \gs~, \\
  \gd \chi_{mn} = H_{mn}~, & \gd H_{mn}=-i\L^A_{\sreeb}\chi_{mn}-[\gs,\chi_{mn}]~, \\
  \gd \gs =-i \reeb^m \Psi_m~, &
\end{array}\label{susy_vect_twist}\eea
 where $\L^A_{\sreeb} = L_{\sreeb} + i [~,\iota_{\sreeb} A] $.
 Here the 2-form $H$ is a redefinition of $D$
\bea
H_{mn}=2(F_H^+)_{mn}+(\xi^I\Gc_{mn}\xi^J)(D_{IJ}+\frac{2}{r}t_{IJ}\gs)~,\label{D_to_H}\eea
where
\bea F_H^+=\frac{1}{2}(1+*_{\sreeb})F-\frac{1}{2}(\kappa\wedge\iota_{\sreeb}F)\nn\eea
is the horizontal self-dual component of $F$. The field $H$ satisfies exactly the same conditions (\ref{horizontal_self_dual}) as $\chi$.

The square of the transformations (\ref{susy_vect_twist}) reads
\bea
  \delta^2 = -iL_{\sreeb} + G_{i(\sigma +\iota_{\sreeb} A )} ~,\label{susy_closure_vect}
 \eea
 where $G_{i(\sigma +\iota_{\sreeb} A )}$ is a gauge transformation with parameter
  $i (\sigma+\iota_{\sreeb} A )$. With our conventions $G_{\ep}$ acts as
\bea
 G_{\epsilon} A = d\epsilon - i [A, \epsilon]~,~~~G_{\epsilon} \phi  = -i [\phi    ,\epsilon ]~,\nn\eea
where $\phi$ is any field in the adjoint.

As promised, (\ref{susy_closure_vect}) shows that the susy transformation written in the new variables can be regarded as an equivariant differential, with the fields $A_m$, $\chi_{mn}$ being the coordinates of some space and $\Psi_m,~H_{mn}$ their conjugate momenta. The field $\gs$ is a bit special, since the combination $\Phi=\gs-\iota_{\sreeb}A$ is annihilated by $\gd$, but this field will also be incorporated as a momentum once the ghost sector is included \cite{Pestun:2007rz}.
\subsection{Hyper-Multiplet}
\label{hyper}

For the hyper-multiplet, we would do what is opposite to the previous section, and combine the scalar field $q_I$ in (\ref{hyper-tran-noaux}) with the Kiling spinors and formulate the cohomoogical complex in terms of spinors (the goal is always to work with a complex that is $SU(2)_R$-singlet).
We define a new bosonic spinor field $q$
\bea q=\xi_Iq^I~,~~~~~q_I=-2\xi_I q~.\nn \eea
From the reality condition satisfied by $\xi_I$ and $q_I$ one can see that the spinor field $q$ now satisfies the same reality condition as $\psi$.

The susy transformation (\ref{hyper-tran-noaux}) expressed in terms of $q$ and $\psi$ will only close on-shell, to mend this we need introduce a bosonic spinor field ${\cal F}$ of opposite $\gc_5=-\reeb\cdotp\Gc$ eigenvalue. One can obtain an off-shell susy enlarging (\ref{hyper-tran-noaux})
  \bea
  && \delta q^A = iP_+\psi^A~,\nn\\
 && \delta \psi^A =-\frac{1}{4r}J_{pq}(\Gc^{pq}q^A)+(\slashed{D}+i\gs) q^A+{\cal F}^A~,\label{susy_hyper_twist}\\
&& \delta {\cal F}^A = -iP_-\slashed{D}\psi^A-\gs P_-\psi^A-\Psi^m(\Gc_m+\reeb_m)q^A~,\nn
\eea
where we use the projector $P_{\pm}=1/2(1\pm \gamma_5)$ and $P_+ q = q$, $P_- {\cal F}={\cal F}$. Notice that $\gs q^A$ should be understood as $\gs^A_{~B}q^B$ and similarly for the term involving $\Psi_m$. The transformations above square to the following:
\bea
\delta^2\Phi=\big(-iL^s_{\sreeb}-\gs-\iota_{\sreeb}A\big)\Phi~,~~~~\Phi=\{q,\psi,{\cal F}\}~,\label{closure_hyper}\eea
where $L_{\sreeb}^s$ is the spinor Lie derivative, see appendix \ref{A-spinors}.
After a further linear shift of ${\cal F}$, one can break the middle line of (\ref{susy_hyper_twist}) according to its eigenvalue under $\gc_5$, and get a nice complex that parallels (\ref{susy_vect_twist})

\begingroup
\renewcommand*{\arraystretch}{1.2}
\bea
&&\begin{array}{ll}
  \delta q^A=i\psi^A_+~, & \gd \psi^A_+=\big(-L_{\sreeb}^s+i(\gs+\iota_{\sreeb}A)\big)q^A~, \\
  \gd\psi^A_-=\tilde {\cal F}^A, & \gd \tilde{\cal F}^A=\big(-iL_{\sreeb}^s-(\gs+\iota_{\sreeb}A)\big)\psi^A_-~.
\end{array}
\label{coho_hyper_spin}\eea
\endgroup
The above complex is written in terms of the fields $q,\psi$ and $\tilde{\cal F}$ which satisfy the reality conditions.
We can solve these reality conditions in terms of the unconstrained fields as we did in (\ref{rewrite_fermion}). Let
\bea q^A=[q^\alpha,-Cq^*_\beta]^T~,~~~~\psi^A=[\psi^\alpha,-C\psi^*_\beta]^T~,~~~~\tilde{\cal F}^A=[\tilde{\cal F}^\alpha,-C\tilde{\cal F}^*_\beta]^T~.\label{Atoalpha}\eea
Now we
can rewrite the complex (\ref{coho_hyper_spin}) in terms of the new fields and it looks exactly the same, except for the change of indices $A\to \alpha$
\bea
 \delta q^\alpha=i\psi_+^\alpha~,~~~\delta\psi_+^\alpha=\cdots~.\nn
\eea
One property that we will need for the transformations is that it acts holomorphically, in that it does not mix $q^\alpha,\psi^\alpha,\tilde{\cal F}^\alpha$ with their conjugates.
This point will be important later when we decide over what spaces do we compute the determinant of the operator $\delta^2$, see next section.

\subsection{Localisation}\label{sec_localisation}
The localisation argument has now become fairly standard, so we will not give all the details. Take a finite dimensional example
\bea \int~ d^nx~d^n\psi ~\go(x,\psi)~,\nn\eea
where $x$ is regarded as the coordinate (even or odd) of certain space and $\psi$ its momentum of the opposite statistics.
Assume that $\go(x,\psi)$ is invariant under an odd symmetry
\bea \gd x=\psi~,~~~~~\gd\psi=L_vx~.\nn\eea

Pick a function $V$ odd, such that $\gd^2V=0$, one can insert into the integral a factor
\bea
\int~ d^nxd^n\psi~\go(x,\psi)e^{-t\gd V}~,~~~~\gd^2V=0~,\nn
\eea
without changing the value of the integral. The last statement can be seen by differentiating with respect to $t$, and using $\int\, \gd(\cdots)=0$. If one then sends $t\to\infty$, the integral will be concentrated at the critical points of the bosonic part of $\gd V$
\bea
\int~ d^nx~d^n\psi~\go(x,\psi)e^{-t\gd V}=\sum_{\textrm{cr pt}}\go~ \textrm{sdet}^{-1/2}\big((\gd V)^{''}\big)~,\nn\eea
where of course we assume that the critical points are non-degenerate.

Furthermore, at each critical point, the equality $\gd^2V=0$ leads to certain relations among the coefficients of Hessian $(\gd V)''$, and consequently the simplification
\bea
\textrm{sdet}^{-1/2}\big((\gd V)^{''}\big)=\textrm{sdet}^{-1/2}_x(\gd^2)~,\nn
\eea
where $\textrm{sdet}_x$ means to take the super-determinant only on the coordinates.

In the case one has complex coordinates $x,\bar x$, complex momenta $\psi,\bar\psi$ and $\gd$ acts holomorphically, then a similar argument as above gives, up to a constant phase, the determinant
\bea \textrm{sdet}^{-1/2}\big((\gd V)^{''}\big)=\textrm{sdet}^{-1}_{hol~x}(\gd^2)~,\nn\eea
where the subscript ${}_{hol~x}$ means taking the super-determinant only on the holomorphic coordinates. Furthermore, one has $\textrm{sdet}_{hol~x}(\gd^2)=\textrm{sdet}_{ahol~x}(\gd^2)$, again up to a constant phase. Note that this phase can be computed for a finite dimensional case, but for an infinite dimensional problem, we have yet no means to handle it systematically, this problem is left open both in \cite{Kallen:2012va} and in the current work.

To summarise, the final result of the integral is just the sum of contribution from each critical point of the above form. The first case above applies to the vector multiplet while the second to the hyper-multiplet.

\subsection{Localisation Locus}
To put this knowledge to practice, we look at the vector multiplet first. One adds to the action an exact term
\bea &&S_{vec}\to S_{vec}+t\gd\int V_{vec}~,\nn\\
&&V_{vec}=\Tr\Big[\frac12\Psi\wedge*(-\iota_{\sreeb}F-D\gs)-\chi\wedge* H+2\chi\wedge * F\Big]~.\nn\eea
The bosonic part of $\delta V_{vec}$ is
\bea
\gd V_{vec}\big|_{bos}&=&\Tr\Big[\frac12(-\iota_{\sreeb}F+D\gs)\wedge*(-\iota_{\sreeb}F-D\gs)-H\wedge* H+2H\wedge* F\Big]\nn\\
&=&\Tr\Big[\frac12(\iota_{\sreeb}F)^2-\frac12(D\gs)^2+F_H^+\wedge* F_H^+-(H-F_H^+)\wedge*(H-F_H^+)\Big]~.\nn\eea
One can integrate out $H$, while the rest is a perfect square\footnote{keep in mind that $\gs$ has been Wick rotated and hence is valued in $i\FR{g}$}, so the localisation locus is
\bea
F_H^+=0~,~~~\iota_{\sreeb} F =0~,~~~D\sigma=0~.\label{general-vect-loc1}
\eea
The first two equations describe the so called 'contact instanton' while the last says $\gs$ is a covariant constant. To keep the discussion lucid, we have not mentioned the gauge fixing ghost sector, but the details can be found in \cite{KallenZabzine12}.

For the geometry $S^5$, the Reeb vector is generated by a free $U(1)$ action and the quotient is $\BB{C}P^2$, one calls these SE manifolds \emph{regular}.
From the analysis of \cite{Kallen:2012va}, the contact instantons on $S^5$ correspond to usual instantons on $\BB{C}P^2$, and only the irreducible ones contribute. The argument leading to this relies on a specific choice of gauge. To reach this gauge one needs to use the regularity condition. The SE manifolds $Y^{p,q}$ are generically \emph{irregular} as the Reeb vector field does not have closed orbit. In the best case, for certain values of $p,q$, when the Reeb generates a locally free action, then the manifolds are a $U(1)$ fibration over a base orbifold. Due to these concerns, it is not clear to us whether the contact instanton on $Y^{p,q}$ can be thought of as usual instantons on the base orbifold in the quasi-regular case, let alone the irregular case. But at any rate, it is likely to be more advantageous to reverse the game and study 4D instantons by lifting them to 5D rather than pushing 5D instantons to 4D. We leave this subject for future enterprise and focus on the zero-instanton sector for the rest of the paper. We do want to point out that the vanishing argument proved in subsection\ref{sec_vanishing} is valid for all contact instanton backgrounds.

\smallskip

\noindent$\sbullet$ For the hyper-multiplet, one can add to the action a $\gd$-exact term
\bea
S_{hyp}\to S_{hyp}+t\gd \int\, V_{hyp}~,~~~~\textrm{where}~~~~V_{hyp}=\frac12(\gd\psi^A)^{\dagger}\psi^A~.\nn
\eea
The bosonic part of $\delta V_{hyp}$ is \footnote{In writing $\gd V_{hyp}$ we used the original complex (\ref{susy_hyper_twist}), in particular we used the field ${\cal F}$ instead of the shifted variable $\tilde{\cal F}$. This is different from what we did in \cite{Kallen:2012va}}
\bea \gd V_{hyp}\big|_{\textrm{bos}}=\big|(-\frac14\slashed{J}+\slashed{D})q\big|^2+\big|\gs q\big|^2+\big|{\cal F}\big|^2~,\nn\eea
where we have already used (\ref{Atoalpha}) to write all fields in terms of their unconstrained components.
In the manipulation above $\gs$ and $\tilde{\cal F}$ are Wick rotated, which is crucial for having a complete square as above.

Since $\gd V$ is positive definite the localization locus is given by the following equations
\bea \big(-\frac14\slashed{J}+\slashed{D}\big)q^\alpha=0~,~~~\gs^\alpha_{~\beta} q^\beta=0~,~~~~{\cal F}^\alpha=0~.\label{locus_hyper}\eea
We first point out that by applying the chiral projector to the first equation, one gets two conditions
\bea P_-\slashed{D}q=0~,~~~\big(-\reeb\cdotp D+\frac14\slashed{J}\big)q=\big(-L_{\sreeb}^s+i\iota_{\sreeb}A\big)q=0~,\label{implied_condition}\eea
in fact if one had used the complex (\ref{coho_hyper_spin}), then $P_-\slashed{D}q$ is absorbed into $\tilde {\cal F}$.

It was shown in \cite{Kallen:2012va} that this set of conditions implies $q={\cal F}=0$ at the $A=0$ configuration.
Now we make a digression and  prove a vanishing theorem that strengthens this result, then we will resume with the partition function.

\subsection{The Vanishing Argument for Hyper-Multiplet}\label{sec_vanishing}

We start from the equation $(\slashed{D}-\slashed{J}/4)q=0$ and show that $q=0$ at an instanton background. Consider the intergral
\bea
0&=&\int\limits_{M}~\Big((-\frac{1}{4}\slashed{J}+\slashed{D})q\Big)^{\dagger}\cdot(-\frac{1}{4}\slashed{J}+\slashed{D})q=\int\limits_M q^{\dagger}(\frac{1}{4}\slashed{J}+\overleftarrow{\slashed{D}})(-\frac{1}{4}\slashed{J}+\slashed{D})q\nn\\
&=&\int\limits_M q^{\dagger}\big(-\slashed{D}^2+\frac{1}{4}\slashed{J}\slashed{D}-\frac14\overleftarrow{\slashed{D}}\slashed{J}-\frac{1}{16}\slashed{J}^2\big)q
=\int\limits_Mq^{\dagger}\big(-\slashed{D}^2+\frac{1}{8}\slashed{J}^2-\frac{1}{16}\slashed{J}^2\big)q\nn\\
&=&\int\limits_M q^{\dagger}\big(\frac{1}{16}\slashed{J}^2-\slashed{D}^2\big)q~.\nn\eea
 The two terms in the integral are
\bea
\slashed{D}^2=D^2-5-\frac{i}{2}\slashed{F}~,~~~~~\slashed{J}^2=-16P_+\nn~.\eea
We also put the gauge field in an instanton configuration $\iota_vF=0=F_H^+$. Then we have
\bea q^{\dagger}\slashed{F}q=q^{\dagger}(F_H^+)_{mn}\Gc^{mn}q=0~.\nn\eea
Assembling everything altogether
\bea
0=\int\limits_M q^{\dagger}\big(\frac{1}{16}\slashed{J}^2-D^2+5\big)q=\int\limits_M q^{\dagger}\big(-D^2+4\big)q=\int\limits_M (D_mq)^{\dagger}(D^mq)+4\int\limits_M q^{\dagger}q~.\nn\eea
So we must have $q=0$.

\vskip.8cm

Now we can write down schematically the perturbative part of the partition function. From the discussion of subsection \ref{sec_localisation}, we first need to evaluate the classical action at the localisation locus. The hyper-multiplet action completely vanishes, while from the vector multiplet action (\ref{action_vector}), we get only
\bea S_{cr}=\int d^5x\sqrt g  \frac{1}{\gYM^2} \Tr\big[-\frac12 D_{IJ}D^{IJ}+\frac{2}{r} \gs t^{IJ}D_{IJ}- \frac{10}{{r}^2}
t^{IJ}t_{IJ}\gs^2\big]\Big|_{D_{IJ}=-2t_{IJ}\gs/r,~\gs=\textrm{const}}~,\nn\eea
where $D_{IJ}=-2t_{IJ}\gs/{\SF r}$ is deduced from the change of variable (\ref{D_to_H}), and that at the critical point $H=F_H^+=0$. Thus we get
\bea S_{cr}=\frac{\textrm{Vol}_{Y^{p,q}}}{\gYM^2} \Tr\big[-\frac{16}{r^2}t^{IJ}t_{IJ}\gs^2\big]
=-\frac{8\textrm{Vol}_{Y^{p,q}}}{\gYM^2r^2}\Tr[\gs^2].\nn\eea
The volume of $Y^{p,q}$ is given in Eq.\ref{volume}.
So we trade $\textrm{Vol}_{Y^{p,q}}$ for $\varrho\textrm{Vol}_{S^5}=\pi^3\varrho r^5$
\bea
S_{cr}=-\frac{8\pi^3 r\varrho}{\gYM^2}\Tr[r^2\gs^2]~.\label{classical_action}
\eea

Putting together also the determinant factor at the localisation locus, we get (keep in mind that $\gs$ is purely imaginary)
\bea
Z_{pert}=\int\limits_{i\FR{g}}d\gs~\exp\big(\frac{8\pi^3 \varrho r}{\gYM^2}\,\Tr[r^2\sigma^2]\big)~\frac{\textrm{sdet}'_{vec}(-iL_{\sreeb}-\gs)^{1/2}}{\textrm{sdet}_{hyp}(-iL_{\sreeb}^s-\gs)}~,\nn\eea
where the super-determinant for the vector multiplet is taken over the sections of the complex
\bea
E:~~0\to \Go^0(M)\to \Go^1_H(M)\to \Go^{2+}_H(M)\to0~,\label{complex_vec}\eea
and $\textrm{sdet}'$ denotes the exclusion of constant modes\footnote{see Appendices C.2 and C.3 in\cite{KallenZabzine12}. We only consider the trivial background $A=0$, so the gauge sector of  \cite{KallenZabzine12} derived for $S^5$ is still applicable.}.
In contrast, the super-determinant for the hyper-multiplet is taken over the sections of the spin bundle
\bea
W_{can}:~~0\to \Go^{0,0}(M)\to \Go^{0,1}_H(M)\to \Go^{0,2}_H(M)\to0~.\nn
\eea
In fact one can directly see that $E\simeq W_{can}\oplus W_{can}^*$, which shows that if one considers the hyper-multiplet in the adjoint, there will be extra cancellation between the vector and hyper-multiplet.

We can do some re-writings of the result, first it is convenient to use the dimensionless combination $x=r\gs$, and factor out (an infinite power of) $r$, and consider the determinant of $(-irL_{\sreeb}^s-x)$ and $(-irL_{\sreeb}-x)$ instead. Secondly, since the function to be integrated is ad-invariant, one can write the integral over $\FR{g}$ as an integral over $\FR{h}$ with a Jacobian factor
\bea Z_{pert}=\frac{1}{|W|}\frac{\textrm{Vol}(G)}{\textrm{Vol}(T)}\int\limits_{i\FR{t}}dx~(\prod_{\gb>0}\bra\gb,x\ket^2)\exp\big(\frac{8\pi^3 r\varrho}{\gYM^2}\,\Tr[x^2]\big)
~\frac{\textrm{sdet}'_{vec}(-irL_{\sreeb}-x)^{1/2}}{\textrm{sdet}_{hyp}(-irL_{\sreeb}^s-x)},\label{pert_part}\eea
where $\gb$ runs over positive roots.

The remainder of the paper is about computing these two determinants, we write down here the result, \emph{discarding all the irrelevant multiplicative constants}. Let $\reeb=[\omega_1,\omega_2,\omega_3,\omega_4]$, assuming of course that $\reeb$ is in the interior of the cone dual to the moment map cone (see subsection \ref{sec_TRtAaSEM})
\bea
&&Z_{pert}=\int_{i\FR{t}}dx~\prod_{\gb>0}\bra\gb,x\ket^2\cdotp\exp\big(\frac{8\pi^3 r\varrho[\reeb]}{\gYM^2}\,\Tr[x^2]\big)
~\frac{P_{vec}}{P_{hyp}}~,\label{finale}\\
&&P_{hyp}={\det}_{\underline{R}}\prod_{i,j,k,l\in\Gl^+}\Big(\big(i\omega_1+j\omega_2+k\omega_3+l\omega_4+\frac12(\omega_1+\omega_2+\omega_3+\omega_4)\big)^2-x^2\Big)~,\nn\\
&&P_{vec}={\det}_{adj}\prod_{i,j,k,l\in\Gl_0^+}\Big(\big(i\omega_1+j\omega_2+k\omega_3+l\omega_4\big)^2-x^2\Big)
\cdotp\prod_{i,j,k,l\in\Gl^+_1}\Big(\big(i\omega_1+j\omega_2+k\omega_3+l\omega_4\big)^2-x^2\Big)^{1/2}~,\nn\eea
where the lattices for the products are defined
\bea&& \Gl^+=\big\{i,j,k,l\in\BB{Z}_{\geq0}\;|\;i(p+q)+j(p-q)=kp+lp\big\}~,\nn\\
&& \Gl_0^+=\big\{i,j,k,l\in\BB{Z}_{>0}\;|\;i(p+q)+j(p-q)=kp+lp\big\}~,\nn\\
&& \Gl_1^+=\Gl^+\backslash\big(\Gl^+_0\cup \{0,0,0,0\}\big)~.\nn\eea
We have also used the volume associated with a general Reeb correspondingly\footnote{Surely, for a general Reeb, one does not have an SE metric and hence no susy a priori, but one may take the cohomological complexes (\ref{susy_vect_twist}) and (\ref{coho_hyper_spin}) as the starting point and compute the partition function.} given in (\ref{volume_equiv}).
If one sets
\bea
\omega_1=0~,~~~~\omega_2=\frac{1}{(p+q)\ell}~,~~~~\omega_3=\omega_4=\frac{3}{2}-\frac{1}{2(p+q)\ell}~,\nn
\eea
one obtains the partition function of the supersymmetric theory.  However one may leave the parameters $(\omega_1, \omega_2, \omega_3, \omega_4)$ unfix and  study how the partition function responds to the $U(1)$ isometries of the geometry.

We can do some more rewriting and relate these infinite products to certain generalisation of the Barne's function, or triple sine function. For $P_{vec}$ factor, we can write ${\det}_{adj}f(x)=f(0)^{\textrm{rk}_G}\prod_{\gb\in\textrm{roots}}f(\bra \gb,x\ket)$, but $f(0)^{\textrm{rk}_G}$ is a (non-zero) multiplicative constant that can be discarded. We will also combine the factor $\prod_{\gb>0}\bra \gb,x\ket^2$ with $P_{vec}$ to get
\bea P_{vec}\prod_{\gb>0}\bra \gb,x\ket^2={\det}_{adj}'S^{\Gl}(x|\omega_1,\omega_2,\omega_3,\omega_4)~,\nn\eea
where ${\det}_{adj}'f(x)$ is short for $\prod_{\gb\in\textrm{roots}}f(\bra \gb,x\ket)$, i.e. the determinant taken in the adjoint representation with the zero weight subspace excluded. The function $S^{\Gl}$ is defined as
\bea
S^{\Gl}(x|  \omega_1, \omega_2, \omega_3, \omega_4)=\prod_{(i,j,k,l) \in \Gl^+}\Big(i \omega_1+j
\omega_2+k\omega_3+l\omega_4 +x\Big) \prod_{(i,j,k,l) \in \Gl^+_0}\Big(i\omega_1+j\omega_2+k\omega_3+l\omega_4 -x\Big)~.\label{S_function}\eea
With this new function, the factor $P_{hyp}$ is just
\bea P_{hyp}={\det}_{\underline{R}}S^{\Gl}(x+\frac12(\omega_1 + \omega_2 + \omega_3 + \omega_4)|\omega_1,\omega_2,\omega_3,\omega_4)~.\nn\eea
 Thus finally we get the answer in the form (\ref{finale_intro}) presented in introduction.
  Since mass $M$ for hypermultiplet can be generated through the standard trick of coupling the hypermultiplet to an auxillary vector multiplet we end up with the general expression (\ref{finale_intro})
   for the massive hypermultiplet.
We will leave the investigation of the function $S^{\Gl}(x|  \omega_1, \omega_2, \omega_3, \omega_4)$  to the future work. Here we will be content with establishing that we do get a sensible matrix model, by computing in section \ref{sec_asymptotic_analysis} the asymptotic behaviour of the products, the result is
\bea&& Z_{pert}=\int\limits_{i\FR{t}}dx~\exp\big(\frac{8\pi^3r\varrho}{\gYM^2}\,\Tr[x^2]\big)\exp(\pi V(x))~,\nn\\
&&V(x)\sim
\Tr_{adj}\Big(|x|\big(\frac{q}{2p}\ell+\frac{3}{4}\varrho\big)-\frac{|x|^3}{6}\varrho\Big) -\Tr_{R}\Big(|x|\big(\frac{q}{2p}\ell-\frac{3}{8}\varrho\big)-\frac{|x|^3}{6}\varrho\big)\Big),\label{asymptotic_potential}\eea
which is similar to the case of $S^5$. There is a limit of how big the hyper-multiplet representation can be, beyond this limit, the potential flips over and the matrix is ill-defined.
If the hyper multiplet is in the adjoint, then the cubic term cancels, and the linear potential has coefficient $9/4\varrho$, we have exact matching with the $S^5$ result up to volume factor.
 If we will look at the large volume limit (analogously to the analysis in \cite{Minahan:2013jwa})  then we reproduce
  the known flat case results.

Let us also set $p=1$ and $q=0$ to get the result for $T^{1,1}$. In particular, one has $\varrho=16/27$ and $q\ell=4/9$, and the potential
tends to
\bea
V(x)\sim \Tr_{adj}\Big(\frac{2}{3}|x|-\frac{8}{81}|x|^3\Big) +\Tr_{R}\Big(\frac{8}{81}|x|^3\Big)~.\nn\eea

\section{Computation of the Partition Function}
\label{sec-computation}

\subsection{Spectrum of $L_{\reeb}^s$}
The computation of the super-determinant requires one to find the spectrum of the operators $L_v$ and $L_v^s$. The former is quite straightforward, while the latter will be shown to be equivalent to the former up to a constant but important shift.

We pick the Killing spinor satisfying $D_X\psi_0=-i/2X\psi_0$ (in fact $\psi_0$ would correspond to a zero form in the canonical spin representation (\ref{can_spin_rep})), which satisfies
\bea \frac{1}{2}(1+iJ)X\psi_0=0~,~~~~\forall X~,\nn\eea
so $\psi_0$ serves as the 'vacuum' and any other form can be constructed by applying the 'raising operators'
\bea \ga^1\wedge \cdots \wedge \ga^p\psi_0\in W_{can}~,~~\textrm{where}~~\ga^i\in\Gc(T^*M)~,~~~\frac{1}{2}(1+iJ)\ga^i=0~,\nn\eea
to the vacuum.

By using the fact that $L_X^s$ preserves $J$ and $\left[L_X^s,\Gc\cdotp\ga\right]=\Gc\cdotp (L_X\alpha)$,
we get
\bea L_X^s(\ga^1\wedge \cdots \wedge \ga^p\psi_0)=L_X(\ga^1\wedge \cdots \wedge\ga^p)\psi_0+\ga^1\wedge \cdots \wedge\ga^pL_X^s\psi_0~,\nn\eea
i.e. up to its action on the vacuum, the operator $L_X^s$ is just the usual Lie-derivative on forms.

The next few steps are a bit technical and is presented in the appendix. It turns out that $L_X^s\psi_0=i/rf_X\psi_0$ with $f_X$ being a real constant, and so $L_X^s$ can be identified as $L_X$ with a numerical shift
\bea L^s_X=L_X+\frac{i}{r}f_X~.\nn\eea
The value of this shift is crucial in many ways, and we have our rule of thumb: assume that $X$ has a zero
(or is decomposable into a sum of commuting Killing vectors $X=\sum u_i$, $[u_i,v]=0$, each with a zero).  At one of its zeros, one can assume that $X$ induces rotations of disjoint 2-planes, \emph{for each such rotation of degree $k$ on a 2-plane with the standard complex structure, one gets a shift of $k/2$}, and the sum of all the shifts gives $f_X$.

As an example, for the case of $S^5$ embedded in $\BB{C}^3$, the Reeb vector field $\reeb=e_1+e_2+e_3$, where $e_i$ is the rotation of the $i^{th}$ factor in $\BB{C}^3$, thus we get 3 shifts of 1/2 and $f_{\reeb}=3/2$ (This shift was obtained in \cite{Kallen:2012va} through more or less brute force). While for squashed $S^5$, the relevant $\reeb$ is decomposed as $v=\sum a_ie_i$, where $a_i$ are the squashing parameters and the shift becomes $f_{\reeb}=1/2\sum a_i$, and this shift plays an important role in \cite{Lockhart:2012vp,Kim:2012qf}.
For our situation we have the decomposition (\ref{decomp_of true_reeb}), and the shift is also $f_{\reeb}=3/2$.

\subsection{Transversally Elliptic Operators}
In this section, we will \emph{temporarily suppress the dimensionful parameter $r$, and also treat $x$ as a c-number}. Both ingredients can be reinstated easily later.
The type of super-determinants we need to compute for the hyper-multiplet is
\bea \textrm{sdet}(-iL_X^s+x)=\frac{\det (-iL_X+f_X+x)\big|_{\psi_-}}{\det (-iL_X+f_X+x)\big|_{q_+}}~,\label{sdet_hyp}\eea
where we have replaced the vector field $v$ with a general isometry $X$, which descends from a vector field acting on $\BB{C}^4$
\bea X=\omega_1 e_1+\omega_2 e_2+\omega_3 e_3+\omega_4 e_4~.\nn\eea
Clearly $X$ commutes with $\reeb$ and $L_XJ=0$. We do so to put extra knobs onto the partition function, and to see how the latter responds to the symmetry of the geometry.

For the hyper-multiplet, the statistics of the fields $q_+$ and $\psi_-$ is determined by their eigen-value under $\gc_5=-\reeb\cdotp \Gc$, thus one can use the operator
\bea
D=P_-(\Gc^iD_i) P_+~,~~~P_{\pm}=\frac{1}{2}(1\pm \gc_5)=\frac{1}{2}(1\mp\reeb\cdotp \Gc)\label{ope_D_serious}
\eea
to keep track of the cancellation between bosons and fermions. That is to say, the operator $D$ sends every non-zero mode of $q$ to a non-zero mode of $\psi$, with \emph{equal $L_X^s$ eigenvalue},
and hence the non-zero modes do not contribute anything to (\ref{sdet_hyp}). The mismatch of
 eigenmodes of $L_X^s$ between $q$ and $\psi$ is then in the kernel and cokernel of $D$ and is captured by the index theorem.

To summarise, we need to obtain the kernel and cokernel of $D$, moreover, we need to decompose the (co)kernel into the simultaneous eigen-modes of $e_i$. Put more formally,
we need to compute the equivariant index
\bea \textrm{ind}_G(D)=\textrm{char}_G(\ker D)-\textrm{char}_G(\textrm{coker}\, D)~,\label{index_hom}\eea
where $G$ is the group generated by the three $U(1)$'s and char${}_G$ denotes the character. In the case $D$ is an elliptic operator, the right hand side of (\ref{index_hom}) is also
known as the \emph{Lefschetz number}, which, upon evaluating the character at $1\in G$, gives the usual index.
But now the key difference is that $D$ is not elliptic, its symbol $\gs(D)$ (obtained by replacing $\partial_m$ with $p_m$ in $D$, where $p_m$ is coordinate for $T^*M$) has a kernel along the Reeb direction.

What we here is a \emph{transversally elliptic operator}. To be more precise, let $G$ be a Lie group acting on $M$ by isometry and $\tau:~\FR{g}\to TM$ be the infinitesimal action. Let
$E,~F$ be two $G$-equivariant vector bundles on $M$ and $D:~E\to F$ be a differential operator that commutes with the $G$-action. Define
\bea
 T^*_GM=\{\ga\in T^*M\,|\,\bra \ga ,\tau(x)\ket=0,~\forall x\in\FR{g}\}~,\nn
 \eea
then $D$ is transversally elliptic if its symbol is invertible on $T^*_GM$ except at the zero section. In our case, since $\reeb$ is a linear combination of $e_i$, so the possible kernel of $D$ is precluded by the above condition for $T_G^*M$.

It is important to remember that the equivariant index defined in (\ref{index_hom})
will not be a function (ad-invariant if $G$ is not abelian) on $G$, but rather a \emph{distribution}. For example, over $S^1$ with the standard $U(1)$-action, the 0 operator is transversally elliptic, and has delta function $\delta(1-t)$ as index, where $t\in U(1)$. Thus in general, the index will have torsions and it will be illegal to evaluate the index at a given point.

In the lectures notes \cite{Ellip_Ope_Cpct_Grp} of Sir Michael Atiyah, the index homomorphism was in principle completely worked out. In the following, we will use two different formulae given in \cite{Ellip_Ope_Cpct_Grp}. The first is quite simple and is applicable only because of the simplicity of the $Y^{p,q}$ geometry, especially since it is a quotient of $S^3\times S^3$.

\subsection{Computing the Index on $Y^{p,q}$-Method I}\label{sec_method_I}
The easiest derivation of index theorem is through the equivariant K-theory. To explain the index calculation, we find ourself in need of making a big digression to review a small portion of the book by Atiyah. The symbol $\gs(D)$ induces a bundle map $\pi^*E\to\pi^*F$ (where $\pi:~T^*_GM\to M$ is the projection), and since $\gs(D)$ is an isomorphism away from the zero section, it gives a complex
\bea 0\to \pi^*E\stackrel{\gs(D)}{\to} \pi^*F\to 0\nn\eea
exact except at the zero section, thus the complex is an element of
\bea K^0_G(T^*_GM)\nn\eea
which by definition consists of stable isomorphism classes of pairs of bundles $[E_1,E_2]$ such that $E_1\simeq E_2$ outside of a compact subset of $T^*_GM$. It is also convenient to choose a $G$-invariant metric and thereby identify $T_G^*M$ with $T_GM$, where the latter consists of tangent vectors perpendicular to the $G$-action.

The \emph{index homomorphism} associates an element of $\gs(D)\in K^0_G(T_GM)$ with the index of $D$. But the crucial point is that it is possible to give a \emph{topological characterisation} of this homomorphism, which then tremendously simplifies the index calculation since manipulating isomorphism classes of vector bundles is much simpler than manipulating the differential operators.

We start from a simple example that could actually carry us a long way in our computation for $Y^{p,q}$. Take a single sphere $S^{2n-1}$ embedded in $\BB{C}^n$ and a group $G=U(1)$ acts on $\BB{C}^n$ with charge vector $[m_1,\cdots, m_n]$, with $m_i>0$. Let $H$ also be a $U(1)$ acting with charge $[1,\cdots,1]$, we remark that
through demanding all $m_i>0$ (of course all of them $<0$ is equally good), the two K-groups $K_G^0(T_GS^{2n-1})$ and $K_G^0(T_HS^{2n-1})$ are isomorphic, because
the bundle $T_GS^{2n-1}$ is isomorphic to $T_HS^{2n-1}$. The latter isomorphism is constructed by simply projecting $T_HS^{2n-1}$ to $T_GS^{2n-1}$; since on the sphere the vector field given by the charge vector $[m_1,\cdots,m_n]$ is nowhere orthogonal to the one with charge vector $[1,\cdots,1]$, and hence the projection has no kernel. On $T_HS^{2n-1}$ there is a $\bar\partial$ symbol which is formed by pulling back the $\bar\partial$ operator defined on the quotient $S^{2n-1}/H\simeq \BB{C}P^{n-1}$. We will call this symbol the horizontal $\bar\partial$-symbol, denoted as $[\bar\partial_H]$, as it is defined on the plane perpendicular to the Reeb vector $H$ of $S^{2n-1}$. This symbol, when regarded as an element of $K^0_G(T_GS^{2n-1})$, has index (proposition 5.4 in \cite{Ellip_Ope_Cpct_Grp})
\bea \textrm{ind}_G[\bar\partial_H]=\left[\prod\frac{1}{1-t^{-m_j}}\right]^--\left[\prod\frac{1}{1-t^{-m_j}}\right]^+~,\nn\eea
where $[~]^{\pm}$ means to expand the content in the brace in positive/negative powers of $t$. For example, let $n=2$ and $m_i=1$
\bea \textrm{ind}_G[\bar\partial_H]=\sum_{k\in\BB{Z}}(k+1)t^{-k}~,\nn\eea
the way to read the result is that the coefficient of $t^{-k}$ is the index
\bea \dim H^0(S^2,{\cal O}(k))-\dim H^1(S^2,{\cal O}(k))~.\nn\eea
It is well-known that $\dim H^0(S^2,{\cal O}(k))=k+1,~k\geq0$ and 0 if $k<0$; while $\dim H^1(S^2,{\cal O}(k))=0,~k\geq0$ and $-1-k$ if $k<0$.
In general, if some $m_i\neq 1$, the index becomes
\bea \textrm{ind}_G[\bar\partial]=\sum_{k,l\geq0}t^{-km_1-lm_2}-\sum_{k,l<0}t^{-km_1-lm_2}~,\label{index_del_bar}\eea
the coefficient of $t^{-m}$ then can be construed as the Riemann-Roch theorem of the weighted projective space (proposition 10.1 in \cite{Ellip_Ope_Cpct_Grp})
\bea \dim H^0(\BB{C}P(\bar m_1,\bar m_2),{\cal O}(m))-\dim H^1(\BB{C}P(\bar m_1,\bar m_2),{\cal O}(m))~,\nn\eea
where $\bar m_1/\bar m_2=m_1/m_2$ and $\gcd(\bar m_1,\bar m_2)=1$. The weighted projective space is an orbifold unless $\bar m_1=\bar m_2=1$.

Back to operator $D$ defined in (\ref{ope_D_serious}). The group in question will be denoted $H=U(1)^3$ (recall that $Y^{p,q}$ is toric).
We need to manipulate $D$ into something manageable. First notice that only the homotopy type of the symbol of $D$ is important for the index computation; second, if one uses the spinor representation by means of the horizontal anti-holomorphic forms, one can unfold the complex $0\to \Go_H^{0,\textrm{even}}\stackrel{D}{\to}\Go_H^{0,\textrm{odd}}\to0$ into
\bea 0\to \Go_H^{0,0}\to \Go_H^{0,1}\to \Go_H^{0,2}\to 0~,\nn\eea
and the symbol $[D]$ turns into the symbol $[\bar\partial_{\reeb}]$ (the operator $\bar\partial$ does not exist, but its symbol does), the subscript ${}_{\reeb}$ is there to remind us that the symbol depends on the horizontal plane and hence $\reeb$. We want to deform $[\bar\partial_{\reeb}]$ into $[\bar\partial_{\reeb_0}]$ that is defined with a more convenient Reeb.
Notice that by looking at the charge vector of $\reeb_0$ and $\reeb$, one easily checks that $\reeb$ and $\reeb_0$ are never anti-parallel (in the dense open subset of $Y^{p,q}$, this is obvious, one needs only pay attention to those points where certain torus action degenerates). One then has an isomorphism between the two complexes $\Go_H^{0,\sbullet} \simeq \Go_{H_0}^{0,\sbullet}$, defined for $\reeb$ and $\reeb_0$ respectively, by simply projecting one to the other. That $\reeb$ and $\reeb_0$ are never anti-parallel guarantees that the projection has no kernel, and hence an isomorphism. Under this isomorphism, the symbol $[\bar\partial_{\reeb}]$ induces a symbol homotopic to $[\bar\partial_{\reeb_0}]$, and we can thus compute the index using the new symbol $[\bar\partial_{\reeb_0}]$. To compute the index, we lift the symbol $[\bar\partial_{\reeb_0}]$ from $Y^{p,q}$ to $S^3\times S^3$, and enlarge the group $H$ to $G=H\times U(1)$, where the extra $U(1)$ is the freely acting $U(1)$ called $U(1)_T$ before. We will lift the symbol to $S^3\times S^3$ through the projection $S^3\times S^3\to Y^{p,q}$. Because this $U(1)_T$ is free, once we have computed the index on $S^3\times S^3$, we can pick out the terms that correspond to the trivial representations of $U(1)_T$ (see theorem 3.1 in \cite{Ellip_Ope_Cpct_Grp}), so as to go back down to the quotient space $Y^{p,q}$. In lifting from $Y^{p,q}$ to $S^3\times S^3$, the horizontal complex $[\bar\partial_{\reeb_0}]$ is lifted to the $\bar\partial$-complex that is horizontal w.r.t both $T$ and $\reeb_0$. To see this, it is useful to keep in mind the following picture. One can obtain the horizontal $\bar\partial$-complex on $Y^{p,q}$ in two steps. One starts from the standard complex structure on $\BB{C}^4$, restricting oneself to the constant moment map level $\mu_T^{-1}(0)$ (where $\mu_T$ is the moment map for $U(1)_T$), one has a transverse complex structure transverse to $T$. This complex structure is clearly invariant under $U(1)_T$ and will go down the symplectic quotient to be the complex structure ${\cal J}$ on $\BB{C}^4//U(1)_T$, and as we recall, the latter is the K\"ahler cone $C(Y^{p,q})$. In the second step, we restrict ${\cal J}$ to the plane transverse to both the homothetic vector field $r\partial_r$ and the Reeb vector ${\cal J}(r\partial_r)$, and we obtain the desired transverse complex structure on $Y^{p,q}$. Since $S^3\times S^3$ is obtained from $\BB{C}^4$ by imposing $\mu_T=0$ and $\sum_i|z_i|^2=1$, we need only restrict ourselves to the directions transverse to $T$ and Reeb to get the last mentioned transverse complex structure.

Having established this, the $\bar\partial_{\reeb_0}$-complex splits into the horizontal $\bar\partial$-complex of the two $S^3$'s individually. What remains is to compute the index of $[\bar\partial]$ on the two $S^3$, take the product, and pick out the terms of trivial representation of $U(1)_T$.
Let now $G_1$ (resp. $G_2)$ be $U(1)^2$ acting on the first (resp. second) $S^3\subset \BB{C}^2$ in the standard manner. Denote by $s_{1,2}$ and $t_{1,2}$ the coordinates of the two $U(1)'s$ of $G_1$ (resp. $G_2$). Then as a simple generalisation from the above example, the index is
\bea &&\textrm{ind}_{G_1}[\bar\partial]=\sum_{i,j\geq0}s_1^{-i}s_2^{-j}-\sum_{i,j<0}s_1^{-i}s_2^{-j}~,\nn\\
&&\textrm{ind}_{G_2}[\bar\partial]=\sum_{k,l\geq0}t_1^{-k}t_2^{-l}-\sum_{k,l<0}t_1^{-k}t_2^{-l}~,\nn\eea
and the index of the product $S^3\times S^3$ is the product of the two indices above. Since $T$ has charge vector $[p+q,p-q,-p,-p]$, the terms in the product of the two indices above that correspond to the trivial representation under $T$ must satisfy
\bea i(p+q)+j(p-q)=(k+l)p~.\label{equal_power}\eea
This is possible only between the following combination
\bea \sum_{i,j\geq0}s_1^{-i}s_2^{-j}\cdotp\sum_{k,l\geq0}t_1^{-k}t_2^{-l}+\sum_{i,j<0}s_1^{-i}s_2^{-j}\cdotp\sum_{k,l<0}t_1^{-k}t_2^{-l}\Big|_{i(p+q)+j(p-q)=(k+l)p}~.\label{index_M_I}\eea
We define a lattice
\bea \Gl^+&=&\{i,j,k,l\geq0,~i(p+q)+j(p-q)=p(k+l)\}~,\nn\\
\Gl^-&=&\{i,j,k,l<0,~i(p+q)+j(p-q)=p(k+l)\}~,\nn
\eea
and our summation of the indices will take place on this lattice.

Let us pause for a minute and understand what we got. It was shown in sec.\ref{sec_GoYpq} that $Y^{p,q}$ is a $U(1)$-fibration over an orbifold base if one uses $\reeb_1$ instead of the irregular $\reeb$. The Reeb $\reeb_1$ is only locally free, and for given $i,j,k,l$ its mode is
$n=(p+q)i+(p-q)j+p(k+l)$. One can try to reorganise the lattice by fixing the level $n$, then the intersection of constant $n$ plane with the lattice $\Gl^{\pm}$ has only finite number of lattice points (this amounts to looking at Riemann-Roch theorem on an orbifold). But the intersection is quite jagged in the sense that there is no general formula for the number of lattice points at each given $n$, this reflects the fact that the Reeb acts non-freely and we have an orbifold base. However, in the case of $T^{1,1}$, the intersection is nice, since $T^{1,1}$ is regular SE and the base of the Reeb fibration is a (K\"ahler-Einstein) manifold. To see this, let us set $p=1,~q=0$, then for fixed $n$, the intersection has exactly $(n+1)^2$ lattice points, which shows again that $T^{1,1}$ is a $U(1)$ bundle over $S^2\times S^2$ of degree 1 and 1.

\begin{figure}[h]
\begin{center}
\begin{tikzpicture}[scale=.8]
\draw [step=0.3,thin,gray!40] (-2.3,-2.3) grid (2.3,2.3);

\draw [->] (-2.5,0) -- (2.5,0) node [below] {\small$i$};
\draw [->] (0,-2.5) -- (0,2.5) node [left] {\small$j$};

\draw [-,blue] (0,1.8) -- (.6,2.4);
\draw [-,blue] (0,.9) -- (1.5,2.4);
\draw [-,blue] (0,0) -- (2.4,2.4);
\draw [-,blue] (0.9,0) -- (2.4,1.5);
\draw [-,blue] (1.8,0) -- (2.4,.6);

\draw [-,blue] (-0.3,-2.1) -- (-.6,-2.4);
\draw [-,blue] (-0.3,-1.2) -- (-1.5,-2.4);
\draw [-,blue] (-0.3,-0.3) -- (-2.4,-2.4);
\draw [-,blue] (-1.2,-0.3) -- (-2.4,-1.5);
\draw [-,blue] (-2.1,-0.3) -- (-2.4,-.6);

\node at (1.5,1.0) {\small$\Gc^+$};
\node at (-1.5,-1.0) {\small$\Gc^-$};
\end{tikzpicture}\caption{The lattice $\Gc$ for summation for $p=3$ and $q=1$}\label{fig_sum_lattice}
\end{center}
\end{figure}
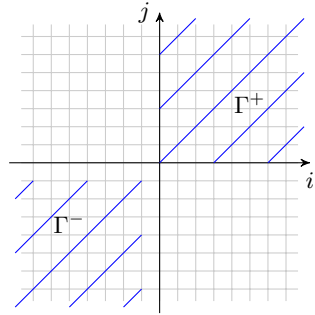

To investigate the lattice $\Gl$ further, we notice that the condition (\ref{equal_power}) is satisfied iff
\bea \Big\{
       \begin{array}{c}
         i+j-(k+l)=mq \\
         i-j=-mp \\
       \end{array}~,\label{only_if}\eea
showing that $i,\,j$ must be on the lattice $\Gc$ in fig.\ref{fig_sum_lattice}, and $k+l$ is determined by $i$ and $j$.
For later use, we point out that the mode $m$ appearing here is (negative) of the mode along the free $U(1)$ called $\ga$ in subsection \ref{sec_LfFUiY}, to see this,
we plug (\ref{only_if}) into the charge vector of the free $U(1)$:
\bea [a,-a-2b,b,b]\to ai-(a+2b)(i+mp)+b(2i+m(p-q))=-m(ap+b(p+q))=-m~.\label{mode_alpha}\eea

\subsection{The Super-Determinant of $L_X^s$}

Now we can finish the calculation of the super-determinant (\ref{sdet_hyp}). From the index (\ref{index_M_I}), a summand such as
\bea s_1^{-i}s_2^{-j}t_1^{-k}t_2^{-l}~,\nn\eea
corresponds to a mode with charge $i,~j,~k,~l$ under $e_1,~e_2,~e_3,~e_4$, thus this mode has $-iL_X$ eigen-value
\bea \omega_1 i+ \omega_2 j+\omega_3 k+\omega_4 l\nn\eea
One then easily writes down the super-determinant as the product
\bea
\textrm{sdet}_{hyp}(-iL_X^s+x)&=&\prod_{\Gl^+}\big(h(i,j,k,l,x)\big)\cdotp\prod_{\Gl^-}\big(h(i,j,k,l,x)\big)~,\nn\\
h(i,j,k,l,x)&=&\omega_1 i+\omega_2 j+\omega_3 k+\omega_4 l+(\omega_1 + \omega_2 + \omega_3 + \omega_4)/2+x~.\nn\eea
where we have the extra shift worked out earlier.
We remark here that whenever we write an infinite product, we mean implicitly the \emph{zeta regulated product} \cite{ZetaProduct}, i.e.
\bea \prod_k\gl_k=\exp\Big(-\frac{\partial}{\partial s}\frac{1}{\Gc(s)}\int_0^{\infty}\sum_ke^{-\gl_kt}t^{s-1}dt\Big|_{s=0}\Big)\label{zeta_reg_prod}\eea
and analytically continuated.

Notice that the product can be put in a form symmetric in $x\to -x$. For the lattice $\Gl^-$, one redefine $i=-i-1$, $j=-j-1$, $k=-k-1$ and $l=-l-1$, then the product can be written as
\bea \prod_{\Gl^-}h(i,j,k,l,x)=\prod_{\Gl^+}h(-i,-j,-k,-l,x)~,\nn\eea
from the rhs one can pull out (an infinite number of) minus signs and rewrite
\bea rhs=(-1)^{\infty}\prod_{\Gl^+}h(i,j,k,l,-x)~.\nn\eea
If one so wishes, one can use the zeta function to regulate also $(-1)^{\infty}$, but as it is a constant, we will just discard it. Finally
the super-determinant is
\bea \textrm{sdet}_{hyp}(-iL_X^s+x)=\prod_{\Gl^+}h(i,j,k,l,x)\cdotp h(i,j,k,l,-x)~.\label{sdet_hyper_grid}\eea
We remark that the symmetry of $x\to -x$ in the result is no coincidence, since the matter content of the hyper-multiplet transforms in the representation $\underline{R}\oplus \underline{\bar R}$ of the gauge group, and $x\to -x$ corresponds to taking $\underline{R}\to \underline{\bar R}$, so here we have a nice confirmation of this symmetry.

For the sake of variation as well as double check, we present in the appendix a different calculation also based on formulae given in \cite{Ellip_Ope_Cpct_Grp}. This method resembles in appearance the Lefschez fixed point formulae (see \cite{AtiyahBottII}) but differs drastically in its interpretation. The computation can be performed on $Y^{p,q}$ without lifting the geometry to $S^3\times S^3$.

We need to put the product into a form more suitable for the asymptotic analysis.
Note that (\ref{only_if}) has two subcases
\bea m\geq0,~i\geq0:~~ j=i+mp,~~k+l=2i+m(p-q)~,\nn\\
m>0,~j\geq0:~~ i=j+mp,~~k+l=2j+m(p+q)~,\nn\eea
the first applies to the part of $\Gc^+$ above (and including) the diagonal while the second below the diagonal. Plugging these into the product
\bea&& \textrm{sdet}_{hyp}(-iL_X^s+x)=\prod_{m,i\geq0}\prod_{k=0}^{2i+m(p-q)}h(i,i+mp,k,2i+m(p-q)-k,x)\cdotp h(\cdots,-x)\nn\\
&&\hspace{3.2cm}\prod_{m>0,j\geq0}\prod_{k=0}^{2j+m(p+q)}h(j+mp,j,k,2j+m(p+q)-k,x)\cdotp h(\cdots, -x)~.\label{sdet_hyper_convenient}\eea

\vskip0.5cm
For the vector multiplet, one can recycle most of the results above by mapping the complex in (\ref{complex_vec}) to $E\simeq W_{can}\oplus W_{can}^*$. As such, the super determinant $\textrm{sdet}_{vec}'(-iL_X+x)$ will be the norm squared of that of the hyper-multiplet, but without the shift and without the constant modes. In total we get
\bea \textrm{sdet}_{vec}'(-iL_X+x)&=&\prod_{\Gl^+\backslash\{0\}}g(i,j,k,l,x)\cdotp\prod_{\Gl^-}g(i,j,k,l,x)~,\nn\\
g(i,j,k,l,x)&=&\big((\go_1i+\go_2j+\go_3k+\go_4l)^2-x^2\big)~.\label{sdet_vec_grid}\eea
One can equally flip the $\Gl^-$ part above to $\Gl^+$. It is convenient to divide $\Gl^+\backslash\{0\}$ into one part
$\Gl^+_0=\Gl^+\cap\{i,j,k,l>0\}$ and another part $\Gl^+_1$ where at least one but not all of $i,j,k,l$ are zero.
Finally the product becomes
\bea \textrm{sdet}_{vec}'(-iL_X+x)=\prod_{\Gl_0^+}g(i,j,k,l,x)^2\cdotp\prod_{\Gl^+_1}g(i,j,k,l,x)~.\nn
\eea

%
%

\section{Asymptotic Analysis of the Partition Function}
\label{sec_asymptotic_analysis}

\subsection{General Formulae}
We can work out a slightly more general formula. Let $x$ be a complex variable with $\re x=0$ and $|\im x|\gg0$ and let $g(t)$ be a function such that
the two integrals below converge.
\bea
\int_1^{\infty}\big|g(t)e^{-xt}\big|dt<\infty~,~~~~~\int_1^{\infty}\big|g(t)'e^{-xt}\big|dt<\infty~.\nn
\eea
From the absolute convergence above, we know that
\bea
I=\int_1^{\infty}g(t)e^{-xt}dt\sim{\cal O}(x^{-1})~.\label{estimate}
\eea
The technique that leads to this estimate is rather standard in asymptotic analysis. Consider
\bea
 I=\int_1^{\infty}g(t)e^{-xt}dt\stackrel{ibp}{=}\frac{1}{x}e^{-x}g(1)+\frac{1}{x}\int_1^{\infty}e^{-xt}g(t)'dt~,\nn
 \eea
it is clear that both terms are of order ${\cal O}(x^{-1})$.

Let $f(t)$ be a function that has Laurent expansion at $t=0$ of the form
\bea
 f(t)=f_nt^{-n}+\cdots +f_1t^{-1}+f_0+{\cal O}(t)=\bar f(t)+{\cal O}(t)~,\label{Laurent}
\eea
and that $f(t)$ tends to constant at infinity. We investigate the asymptotic behaviour of the following
\bea
 J_0^{\infty}(f)=\partial_s\Big(\Gc(s)^{-1}\int_0^{\infty}f(t)e^{-xt}t^{s-1}dt\Big)\Big|_{s=0}~,\nn
 \eea
From the estimate (\ref{estimate}), as well as the fact that $\Gc(0)^{-1}=0$ we will be free to alter the limits of the integral as it suits us
\bea
J_0^{\infty}(f)=J_0^1(f)+{\cal O}(x^{-1})=J_0^1(\bar f)+{\cal O}(x^{-1})=J_0^{\infty}(\bar f)+{\cal O}(x^{-1})~,\nn
\eea
where $\bar f$ is defined in (\ref{Laurent}). It is easy to see
\bea J_0^{\infty}(t^{-k})=\frac{\partial}{\partial s}\frac{\Gc(s-k)}{\Gc(s)}x^{-s+k}\Big|_{s=0}=\frac{\partial}{\partial s}\frac{1}{(s-k)\cdots(s-1)}x^{-s+k}\Big|_{s=0}\nn\eea
for some small $k\geq0$ we have\footnote{where $\ln x$ denotes $\log x$ in its principle branch i.e. $-\pi<\im\ln x\leq\pi$}
\bea J_0^{\infty}(t^{-k}):&&k=0,~~-\ln x;~~~k=1,~~x\ln x-x;\nn\\
&&k=2,~~-\frac12 x^2\ln x+\frac34 x^2;~~~k=3,~~\frac16x^3\ln x-\frac{11}{36}x^3~.\label{form_J}\eea
From these and the Laurent expansion (\ref{Laurent}) one can completely determine the asymptotic behaviour of $J_0^{\infty}(f(t))$.

\subsection{Application to $Y^{p,q}$}

For the hyper-multiplet, we need to work out the asymptotic behaviour of a product of the form (the first line of (\ref{sdet_hyper_convenient}))
\bea
 \star=\prod_{m,i\geq0}\prod_{k=0}^{2i+m(p-q)}h(i,i+mp,k,2i+m(p-q)-k,\pm x)~.\nn
 \eea
From the definition of zeta-regulated product, the relevant integral to consider is
\bea
 -\ln \star&=&\partial_s\frac{1}{\Gc(s)}\sum_{m,i\geq0}\sum_{k=0}^{2i+m(p-q)}\int_0^{\infty}e^{-h(i,j,k,l,\pm x)t}t^{s-1}dt\Big|_{s=0}\nn\\
&=&\partial_s\frac{1}{\Gc(s)}\int_0^{\infty}e^{-(\frac12(\go_1+\go_2+\go_3+\go_4)\pm x)t}t^{s-1}dt\cdotp\frac{1}{1-e^{-(\go_3-\go_4)t}}\nn\\
&&\Big(\frac{1}{1-e^{-(\go_1+\go_2+2\go_4)t}}\frac{1}{1-e^{-(\go_2p+\go_4(p-q))t}}-\frac{1}{1-e^{-(\go_1+\go_2+2\go_3)t}}\frac{e^{-(\go_3-\go_4)t}}{1-e^{-(\go_2p+\go_3(p-q))t}}\Big)\Big|_{s=0}~.\nn\eea
In performing the summation over $i$ and $m$, one needs $\go_1+\go_2+2\go_{3,4}>0$ and $\go_2p+\go_{3,4}(p-q)>0$, but these are implied by the dual cone condition (\ref{dual_cone_condition}).

The second line of (\ref{sdet_hyper_convenient}) can be done similarly. It also follows from (\ref{dual_cone_condition}) that $\sum\go_i>0$, and then the function after the $\cdotp$
satisfies the criteria laid out before and will serve as our $f(t)$. We can expand $f(t)$ into a Laurent series, and
follow the procedure given previously to get the asymptotic behaviour. This involves a bit of meticulous calculation, so we only give the results
\bea
-\ln \textrm{sdet}_{hyp}&=&i\pi\sgn(\im x)\big(\frac16\varrho[\reeb]\cdotp x^3+B_h\cdotp x\big)~,\label{asympt-hyper-general}\\
B_h&=&-\frac{1}{24}\big((\go_3-\go_4)^2+(\go_1-\go_2)^2\big)\varrho[\reeb]\nn\\
&&-\frac{p^2q(\go_1-\go_2)(\go_3\go_4-\go_1\go_2)-pq^2(\go_1\go_2(\go_3+\go_4)+(\go_1+\go_2)\go_3\go_4)}{6((\go_1+\go_3)p+q\go_3)((\go_1+\go_4)p+\go_4q)((\go_2+\go_3)p-q\go_3)((\go_2+\go_4)p-\go_4q)}~.\nn
\eea
 That the coefficient of $x^3$ term is proportional to the general squashed volume is perhaps not surprising. The coefficient $B_h$ satisfies the condition (\ref{constraint}) as expected, but only when taking the two terms together. The geometrical interpretation of $B_h$ is beyond us for now, but it may be related to some correction coming from the non-trivial topology, such as the volume of the generator of $H^2(Y^{p,q})\sim\BB{Z}$. It will be extremely interesting if one can fix this. If we plug in the actual value of $\go_i$ for the Reeb vector (\ref{decomp_of true_reeb}), the result is
\bea
 i\pi\sgn(\im x)\Big(
x\big(\frac{q}{2p}\ell-\frac{3}{8}\varrho\big)+\frac{1}{6}\varrho x^3\Big)~.\nn
\eea

One can also allow $x$ to have a real part, but it must be in the range
\bea \re x\in \big[0,\frac{1}{2}(\go_1+\go_2+\go_3+\go_4)\big),\nn\eea
to make sure both $(\go_1+\go_2+\go_3+\go_4)/2\pm x$ have a positive real part so as to converge the integral. We remark that, in the same computation on $S^5$, the product gives the Barne's function and can be explicitly written in terms of poly-logarithms, which do exhibit branching behaviour when $\re x$ exceeds certain range. Here our function is a generalisation of the Barne's function, and it shows similar branching behaviours.

\vskip.5cm
The calculation for the vector-multiplet is far less pleasant as it lacks the elegant symmetry possessed by the hyper-multiplet, hence we will be just giving the result. But we do point out that the exclusion of the constant modes in the vector part is crucial for the convergence of the integral.
%
\bea -\frac12\ln\textrm{sdet}'_{vec}&=&
\frac12\ln(-x^2)+i\pi\sgn(\im x)\big(\frac{1}{6}\varrho[\reeb]\cdotp x^3+B_v\cdotp x\big),\label{asymp_vec}\\
B_v&=&\frac{1}{12}(\go_3^2+\go_4^2+4\go_3\go_4+3(\go_1+\go_2)(\go_3+\go_4)+\go_1^2+4\go_1\go_2+\go_2^2)\varrho[\reeb]\nn\\
&&-\frac{p^2q(\go_1-\go_2)(\go_3\go_4-\go_1\go_2)-pq^2(\go_1\go_2(\go_3+\go_4)+(\go_1+\go_2)\go_3\go_4)}{6((\go_1+\go_3)p+q\go_3)((\go_1+\go_4)p+\go_4q)((\go_2+\go_3)p-q\go_3)((\go_2+\go_4)p-\go_4q)}~.\nn \eea
If we plug in the value of $\go_i$ for the Reeb again, we get
\bea
\frac12\ln(-x^2)+i\pi\sgn(\im x)\Big(x\big(\frac{q}{2p}\ell+\frac{3}{4}\varrho\big)+\frac{1}{6}\varrho x^3\Big)~.\nn\eea
It is an interesting feature that if one considers a single hyper-multiplet in the adjoint representation, then the leading $x^3$ term cancel just as in the $S^5$ case and the asymptotic behaviour of the potential for the matrix model will be $|x|$. In the next section we will discuss the implication
 of this fact.

\section{$N^3$-behavior from Matrix Model}
\label{sec-matrix}

In this section we present one of the possible applications of our general result (\ref{finale_intro}).
  Let us consider the gauge group $SU(N)$ and the matter content consisting of a single
   hypermultiplet in adjoint representation with mass $M$.
    We are interested in the large $N$ behaviour of the free
    energy for this model. Our analysis is completely analogous to the treatment of the model on $S^5$ from \cite{Kallen:2012zn, Minahan:2013jwa}. Thus we present the final result and for the details we refer
     the readers to the references just given.

 Let us introduce 't Hooft coupling constant $\lambda= g_{YM}^2 N/r$ and
  rewrite the matrix model (\ref{finale_intro}) in terms of eigenvalues $\phi_i$.
In the limit $\lambda \gg 1$ we can assume the large separation of eigenvalues $|\phi_i - \phi_j| \gg 1$
 and moreover we also assume $|M| \ll \lambda $. Using the asymptotic expansions
  (\ref{asympt-hyper-general}) and (\ref{asymp_vec})  the matrix model is drastically simplified in
 the large 't Hooft coupling limit
 \bea
   Z= \int \prod_i d\phi_i ~\exp\Big(-\frac{8\pi^3 N}{\lambda} \varrho \sum\limits_i \phi_i^2 +\frac{\pi}{2} \left [\frac{1}{4}(\sum\limits_{i=1}^{4} \go_i)^2 + M^2 \right ]\varrho  \sum\limits_{i,j} |\phi_i - \phi_j| \Big)~,\nn
 \eea
 where the matrix model is written in terms of $\phi$ eigenvalues.
  Following the same logic as in \cite{Kallen:2012zn}
  we can evaluate the free energy on the saddle point and obtain the following expression
  \bea
   F = - \log Z  = - \frac{g_{YM}^2 N^3}{96 \pi r} \varrho \left ( \frac{1}{4} (\sum\limits_{i=1}^{4} \go_i)^2 + M^2 \right )^2~.
  \eea
 for the general squashed $Y^{p,q}$ space.
 If we consider the case
  of unsquahed $Y^{p,q}$ space  and set $\sum\go_i=3$, in fact, it was shown in \cite{Martelli:2005tp} that $\sum\go_i=3$ is a necessary condition for the Reeb vector to admit an SE metric, then
   we arrive to
   \bea
   F = - \log Z  = - \frac{g_{YM}^2 N^3}{96 \pi r} \varrho \left ( \frac{9}{4}   + M^2 \right )^2~.
  \eea
  Surprisingly the result is identical to that of the theory on $S^5$ up to a volume factor $\varrho$.

 Using the results presented in this paper, it is straightforward exercise to generalise to squashed $Y^{p,q}$ space
  the treatment of matrix models for 5D SCFTs presented in \cite{Jafferis:2012iv}.

\section{Discussion}
\label{sec_discussion}

In this paper we considered the 5D Yang-Mills theory with matter on $Y^{p,q}$ manifolds. This theory preserves two supersymmetries
 and this is sufficient for us to localise the model. The partition function is localised on contact instantons, however, a general treatment of the instanton sector
 presents some challenges due to complications in geometry and is left for the future. We perform explicit
   calculation only for the zero instanton sector and obtained the full perturbative result in terms of certain special functions and we studied their asymptotic properties.

   Let us briefly discuss some related topics:\\

\noindent $\sbullet$ {\bf Isometry and Enlargement of Supersymmetry}\\
We can wonder whether we can enlarge the number of susy  for a given $Y^{p,q}$ geometry. The answer seems to be a disappointing 'no', at least when one does not turn on extra background fields.

Let us investigate the extreme case of $S^5$ embedded in $\BB{R}^6$, which has an $SO(6)$ isometry. Choose three $SO(2)$-rotations along the 1-2, 3-4 and 5-6 direction as the Cartan of $\FR{so}(6)$. The Reeb vector is the sum $\reeb=\sum e_i$, let $X$ be an isometry that does not commute with $\reeb$, and let $L_X^s$ act on the Killing spinors. Since $L^s_X$ preserves the Killing spinors, then $L_X^s\psi$ must also be a Killing spinor linearly independent of $\psi$, this way we have enlarged the number of super charges (this method of analyzing the susy algebra is used in \cite{Moroianu200063} and \cite{FigueroaO'Farrill:1999va}).
Indeed for $S^5$ with the round metric, we get 8 super-charges transforming in $\underline{4}\oplus\underline{\bar 4}$ of $\FR{so}(6)$. For $S^5$ with various squashed metric, one has reduced isometry and hence only a fraction of susy is preserved \cite{Imamura:2012xg}.

For the $Y^{p,q}$ manifold, the infinitesimal isometry is shown to be $\FR{su}(2)\oplus \FR{u}(1)\oplus \FR{u}(1)$ \cite{Gauntlett:2004yd}, but unfortunately all these isometries commute with the Reeb and hence generate no new susy. Actually this follows from a deeper reason.
B\"ar's cone construction \cite{cone_construction} shows that the Killing spinors on $M$ correspond to parallel spinors on the metric cone $C(M)$, and thereby converting the classification problem of Killing spinors to the classification of holonomy on $C(M)$. The latter problem is by now well-understood and the situation is summarised in theorem 5.15 from \cite{Boyer:1998sf}, which shows that $S^5$ is the only 5D manifold with Killing spinors of type (4,4) (this means 4 Killing spinors with $\mu=1$ and 4 with $\mu=-1$), the other SE manifolds admit only Killing spinors of type (1,1).

\vskip.3cm

\noindent $\sbullet$ {\bf Generalisation to $L^{a,b,c}$ spaces}\\
 The spaces $L^{a,b,c}$ contain $Y^{p,q}$ as a subclass (see \cite{Martelli2005208}, whereas the SE metric is presented in \cite{Cvetic:2005ft}). The construction of such spaces is similar, one takes the quotient of $\BB{C}^4$ with respect to a $U(1)$ of charge $[a,b,-c,-a-b+c]$, with appropriate coprimeness conditions on the 3 positive integers $a,b,c$, to form the K\"ahler cone over the desired SE space. Hence, we expect that most of our calculation can be generalised to $L^{a,b,c}$ spaces rather easily, and one obtains the partition function written in terms of the function $S^{\Gl}$, but with the lattice determined by the new charge vector $[a,b,-c,-a-b+c]$. The result may reveal more how the partition function depends on the non-trivial homology of the space.

\bigskip
{\bf Acknowledgements} We thank Johan K\"all\'en, Joseph Minahan, Anton Nedelin and
Konstantin Zarembo for discussion of this and related topics.
The research of J.Q. is supported by the Luxembourg FNR grant PDR 2011-2, and by the UL grant GeoAlgPhys 2011-2013. The research of M.Z. is supported in part by Vetenskapsr\r{a}det under grant $\sharp$ 2011-5079.

\appendix
\section{Notations and Conventions}\label{A-spinors}

We follow the convention for spinors from \cite{HosomichiSeongTerashima}. The gamma matrices satisfy the Clifford algebra
\bea
\{\Gc^{\tt a},\Gc^{\tt b}\}=2\delta^{\tt ab}~,\nn
\eea
and the charge conjugation matrix satisfies
\bea
C^{-1}(\Gc^{\tt a})^TC=\Gc^{\tt a}~,~~~~C^T=-C,~C^*=C~.\nn
\eea
Note that the type-writer fonts are reserved for \emph{flat indices}.

The spinor bi-linears are formed using $C$,
\bea
\psi^TC\chi\stackrel{\textrm{abbreviate}}{\longrightarrow}\psi\chi~,\label{spinor_bi_linear}
\eea
though throughout the paper, these bi-linears are abbreviated as $(\psi\chi)$, following the notation of \cite{HosomichiSeongTerashima}. Due to the symmetry property of $C$, one has
\bea
(\psi\chi)=-(\chi\psi),~~(\psi\Gc^{\tt a}\chi)=-(\chi\Gc^{\tt a}\psi)~,~~(\psi\Gc^{\tt ab}\chi)=(\chi\Gc^{\tt ab}\psi)~,\nn
\eea
where $\Gc^{\tt a_1\cdots a_n}=(1/n!)\Gc^{[\tt a_1}\cdots \Gc^{\tt a_n]}$ and all spinors appearing above are \emph{bosonic} (even). The product of three or more gamma matrices can be reduced
\bea
 \Gc^{\tt abc}e^{\tt abcde}=-6\Gc^{\tt de}~,\label{duality_flat}
 \eea
where $e^{\tt a...b}$ is the Levi-Civita symbol $e^{12345}=1$.

On a curved manifold, one defines the gamma matrices by means of the veilbeins, i.e.\ a set of mutually orthogonal (local) sections of the tangent bundle
\bea E^{\tt a}\in \Gc(TM)~,~~~\bra E^{\tt a},E^{\tt b}\ket=\delta^{\tt ab}~,\nn
\eea
where $\bra-,-\ket$ is the pairing using the metric $g$.
 The gamma matrices are defined as
\bea
\Gc^m=E^{m\tt a}\Gc^{\tt a}~,~~~\Gc_m=g_{mn}\Gc^n~,\nn
\eea
and the duality (\ref{duality_flat}) turns into
\bea
\frac{1}{3!}g^{1/2}\Gc_{mnp}\epsilon^{mnp}_{~~~~\;qr}=-\Gc_{qr}~.\label{duality_curv}
\eea
One may also consult the appendix of \cite{Kallen:2012va} for the Fierz identities.

We also use Dirac's slash notation
\bea \slashed{M}=M_{i_1\cdots i_p}\Gc^{i_1\cdots i_p}~,~~M\in \Go^p(M)~,\nn\eea
and to keep the formulae neat, \emph{we will even dispense with the slash whenever confusion is unlikely}.

%

The $SU(2)$ R-symmetry index are raised and lowered from the left
\bea
\xi^I=\epsilon^{IJ}\xi_J~,~~~~ \xi_I=\epsilon_{IJ}\xi^J~,~~~~ \epsilon^{IK}\epsilon_{KJ}=\delta^I_J~,~~~~ \epsilon^{12}=-\epsilon_{12}=1~.\nn
\eea

\noindent$\sbullet$ The spinor Lie derivative. This notion is quite old \cite{Kosmann}, but it is explained more transparently in the physics context by Figueroa-O'Farrill in \cite{FigueroaO'Farrill:1999va}. For a Killing vector, one defines
\bea
L_X^s=D_X+\frac18\nabla_{[m}X_{n]}\Gc^{mn}=D_X+\frac18\slashed{dX}~,\nn
\eea
which satisfies ($f$ is a function and $Y$ is any vector field)
\bea
&& [L_X^s,\slashed{Y}]=[X,Y]\cdotp \Gc~,~~~~~[L_X^s,f]=L_Xf~,\nn\\
&&[L_X^s,L_Y^s]=L_{[X,Y]}^s~,~~~~~[L_X^s,D_Y]=D_{[X,Y]}~.\label{property_spin_Lie}
\eea
In this paper almost all covariant derivatives will be denoted as $D$, be it on the spin bundle or the gauge bundle; but $\nabla$ will be reserved for the Levi-Civita covariant derivative. The vector indices are raised and lowered with the metric tacitly, e.g. the $dX$ above means identifying $X$ with a 1-form using the metric.

\section{The Shift}

The simple task to accomplish here is to compute $L_X^s\psi$ where $X$ is a Killing vector commuting with $\reeb$ and $\psi$ is a Killing spinor.
We just outline the procedure here.

First it is easy to see that $L_X^s\psi$ will remain a Killing spinor even if $X$ does not commute with $\reeb$, by using the last property of (\ref{property_spin_Lie}). But
we need to show something stronger, i.e. that $L_X^s$ preserves the subbundle defined in (\ref{subbundle}). The first condition of (\ref{subbundle}) is easy, since
\bea [L_X^s,\reeb]\psi=(L_X\reeb)\psi=0~,\nn\eea
using $L_X\reeb=0$. For the second condition, one needs some computation. The following relation will also be useful
\bea 0=L_XJ^p_{~q}&=&\nabla_XJ^p_{~q}-J^r_{~q}\nabla_rX^p+J^p_{~r}\nabla_qX^r\nn\\
  &=&-\reeb^pX_q+X^p\reeb_q-\frac12(dX)_r^{~p}J^r_{~q}+\frac12J^p_{~r}(dX)_q^{~r}\label{dX(2,0)}~,\eea
which places a restraint on the (2,0)+(0,2) part of $dX$.

One then writes
\bea [L_X^s,A(Y)]=[\tilde D_X-\frac{i\mu}{2}X+\frac18\slashed{dX},A(Y)]~,\nn\eea
we already know that $[\tilde D_X,A(Y)]$ restricts to zero. The two latter terms are computed to be
\bea -\frac{i\mu}{2}[X,A(Y)]&=&-i\mu A(Y)+i(XJY)+\frac{\mu}{2}\big((X\cdotp Y)-XY{\reeb}+2X({\reeb}\cdotp Y)-(X\cdotp {\reeb})Y+{\reeb}(X\cdotp Y)\big)~,\nn\\
\big[\slashed{dX},A(Y)\big]&=&-8\mu\iota_Y({\reeb}\wedge X)-4A(\iota_YdX)-4i\mu YA(X)+2\mu Y(1+\reeb)X+8i(YJX)~.\nn\eea
Combine the two pieces, ignoring anything containing $A$ and using ${\reeb}\cdotp\psi=-\psi$, one gets $[L_X^s,A(Y)]=0$. So the second condition of (\ref{subbundle}) is preserved by $L_X^s$.

Since we know that $L_X^s\psi$ remains a Killing spinor, furthermore the subbundle (\ref{subbundle})
 is of rank one, one then sees that $L_X^s\psi$ is a constant multiple of $\psi$.
To find this proportionality coefficient, we look at $L_X^s\psi$ more closely
\bea
 L_X^s\psi=\Big(-\frac{i\mu}{2}X^{||}+\frac{1}{8}(dX^{\perp})_{pq}\Gc^{pq}\Big)\psi~,\nn
\eea
where $X^{||}=({\reeb}\cdot X){\reeb}$ and $dX^{\perp}$ is the horizontal part of $dX$.
Now consider ${dX^{\perp}}$, it can be decomposed into self and anti-self dual part, the latter vanishes when acting on $\psi$, as was shown in (\ref{aselfdualpsi}).
As for the self-dual part, it can be either $(1,1)$ or $(2,0)+(0,2)$ w.r.t $J$, the latter again vanishes due to (\ref{dX(2,0)}). What remains is the self-dual (1,1) part of $dX^{\perp}$, but note that such a 2-form is necessarily proportional to $J$. To see this assume $M$ is (1,1) self-dual and choose the volume form to be $1/2J\wedge J$
\bea
&&M=*_4M=\frac{1}{2}\bra M,J\ket J+JMJ=\frac{1}{2}\bra M,J\ket J+JMJ=\frac{1}{2}\bra M,J\ket J-M~,~~~\To~~~M=\frac14\bra M,J\ket J~,\nn\eea
where $\bra M,J\ket=M_{pq}J^{pq}$.

Thus $L_X^s$ acting on $\psi$ simplifies to a multiplicative factor
\bea L_X^s\psi=\big(\frac{i\mu}{2}(X\cdotp {\reeb})+\frac{1}{32}\bra dX,J\ket \slashed{J}\big)\psi
=i\mu\big(\frac{1}{2}(X\cdotp {\reeb})-\frac{1}{8}\bra dX,J\ket\big)\psi=i\mu f_X\psi\label{L_X^spsi}~,\eea
where we used (\ref{spin_form_degree}).



As a check that $f_X$ is a constant, let us look at $X={\reeb}$ and compute $f_{\reeb}$ from (\ref{L_X^spsi})
\bea f_{\reeb}=\frac12\bra {\reeb},{\reeb}\ket-\frac{1}{8}(-2)\bra J,J\ket=\frac12+\frac14(+4)=\frac32~,\nn\eea
which is the correct result from our $S^5$ computation \cite{Kallen:2012va}. It is nonetheless worth the effort to figure out $f_X$ for a more general $X$, which then allows us to get the fully equivariant partition function.

Now that we know $f_X$ is a constant, we can evaluate it at a convenient point, say, a point where $X=0$, and (\ref{L_X^spsi}) simplifies to
\bea
f_X=-\frac14(\partial_pX^q)J^p_{~q}~,\nn
\eea
where the covariant derivative has been swapped for ordinary derivative since $X=0$. For our particular case, we know $Y^{p,q}$ comes from reduction data, and the complex structure is inherited from the canonical one on $\BB{C}^4$, we can compute the above expression in $\BB{C}^4$.
For each $e_i$, go to one of its fixed point, one has
\bea
(\partial_pe_i^q)J^p_{~q}=-2~,\nn
\eea
the computation is valid passing down to the quotients, since $\iota_T\eta=\mu_T=0$. As a final check, let us plug in the Reeb vector decomposed into $U(1)$'s, namely
\bea \reeb=\big(\frac{3}{2}-\frac{1}{2(p+q)\ell}\big)(e_3+e_4)+\frac{1}{(p+q)\ell}e_2~,\nn\eea
since each $e_i$ contributes $1/2$, we get once again $3/2$.

\section{The Explicit Metric}\label{sec_tEM}
The metric is given by
\bea &&ds^2=\frac{1-cy}{6}{\color{black}(d\gt^2+\sin^2\gt d\phi^2)}+\frac{1}{w(y)q(y)}dy^2+\frac{q(y)}{9}{\color{black}(d\psi-\cos\gt d\phi)^2}\nn\\
&&\hspace{2cm}+w(y)\Big[d\ga+f(y){\color{black}(d\psi-\cos\gt d\phi)}\Big]^2~,\label{Sasaki-Einstein_metric}\\
&&w(y)=\frac{2(a-y^2)}{1-cy}~,~~~~q(y)=\frac{a-3y^2+2cy^3}{a-y^2}~,~~~~f(y)=\frac{ac-2y+y^2c}{6(a-y^2)}~.\nn\eea
The parameter $c$ will be set to be 1 when it is not zero. The angle variable $\ga$ has period $2\pi\ell$, where $\ell$ is defined in (\ref{def_ell}). The parameter $a$ is given by
\bea a=\frac12+\frac14(-1+3\gl^{-2})\sqrt{4-3\gl^{-2}},~~~\gl=p/q~,\nn\eea
and $y$ is restricted to the range $y_1\leq y\leq y_2$, where $y_{1,2}$ is the negative and the smaller positive root of the cubic equation $a-3y^2+2y^3=0$. In particular
\bea
y_1=\frac34(1-\gl^{-1})-\frac{y_3}{2}~,~~~~y_2=\frac34(1+\gl^{-1})-\frac{y_3}{2}~,~~~~y_3=\frac{3}{2}(1-\gl^{-2})+\frac{1}{2p\gl\ell}~.\nn\eea
 The following is easy to check, and will be important later
\bea
 f(y_1)=\frac{\ell}{2}(p+q)~,~~~~f(y_2)=\frac{\ell}{2}(q-p)~.\label{alg_relation}
 \eea
With these choice of parameters, the manifold is homeomorphic to a $U(1)$ bundle over $S^2\times S^2$ with degree $p$ and $q$.

Next, we will try to read off the change of coordinates from the metric, in particular, we want to reproduce table (\ref{equi_action}). The first term is the metric of the round sphere, which is what we call the base sphere. The second and third term describes the fibre sphere, note that the term $-\cos\gt d\phi$ in the combination $d\psi-\cos\gt d\phi$ is the connection for the fibration. In this original way of presenting the metric, it may seem that $d\phi$ is ill-defined at $\gt=0,\pi$, since at the two poles the azimuth angle is undetermined. The reason for this is that the fibration associated to the degree $-2$ $U(1)$ bundle is trivialised everywhere except at the two poles, where each pole hosts a $-1$ point charge (note that $dd\phi$ is not zero but a delta function). In this paper, we prefer to present the metric in the traditional way, namely, we will remove the delta function charge but re-introduce the transition function at the equator. To do so, in the patch $U_{00}:~\gt<\pi,~y>y_1$, redefine $\psi_{00}=\psi-\phi$ and $\phi_{00}=\phi$, but at the patch $U_{10}:~\gt<\pi,~y<y_2$, $\psi_{10}=-\psi+\phi$ and $\phi_{10}=\phi$. This way, in $U_{10}$, the combination
\bea d\psi-\cos\gt d\phi=d\psi_{00}+(1-\cos\gt)d\phi_{00}~,\nn\eea
and similarly for $U_{10}$:
\bea d\psi-\cos\gt d\phi=-d\psi_{10}+(1-\cos\gt)d\phi_{10}~.\nn\eea
But on the intersection $U_{00}\cap U_{10}$, we have $\psi_{10}=-\psi_{00}$.

One has to do the same for the $U(1)$ fibre parameterised by $\ga$. In the patches $U_{00}$ and $U_{10}$, define $\ga_{00}=\ga+f_2\psi_{00}$, and $\ga_{10}=\ga-f_1\psi_{10}$, then
\bea d\ga+f(y)(d\psi-\cos\gt d\phi)&=&d\ga_{00}-f_2d\psi_{00}+f(y)(d\psi_{00}+(1-\cos\gt) d\phi_{00})\nn\\
&=&d\ga_{10}+f_1d\psi_{10}+f(y)(-d\psi_{10}+(1-\cos\gt) d\phi_{10})~,\nn\eea
then the singularity at $y=y_{1,2}$ is removed. On the intersection $U_{10}\cap U_{00}$,
\bea \phi_{10}=\phi_{00}~,~~~~\psi_{10}=-\psi_{00},~~\ga_{10}=\ga_{00}+(f_1-f_2)\psi_{00}~.\nn\eea

Now we get go to the patches covering $\gt=\pi$,
\bea U_{01}:~\{\gt>0,~y>y_1\}~,&&\phi_{01}=-\phi_{00},~~\psi_{01}=\psi+\phi=\psi_{00}+2\phi_{00},\nn\\
&&\ga_{01}=\ga+f_2(\psi+\phi)=\ga_{00}+2f_2\phi_{00}~,\nn\\
U_{11}:~\{\gt>0,~y<y_2\}~,&&\phi_{11}=-\phi_{00},~~\psi_{11}=-\phi-\psi=-2\phi_{00}-\psi_{00}~,\nn\\
&&\ga_{11}=\ga+f_1(\phi+\psi)=\ga_{00}+(f_1-f_2)\psi_{00}+2f_1\phi_{00}~.\nn\eea
From these change of coordinates we get the table below
\bea \begin{array}{c|ccc}
         & U_{10} & U_{01} & U_{11} \\
       \hline
       \partial_{\phi_{00}} & \partial_{\phi_{10}} & -\partial_{\phi_{01}}+2\partial_{\psi_{01}}+2f_2\partial_{\ga_{01}} & -\partial_{\phi_{11}}-2\partial_{\psi_{11}}+2f_1\partial_{\ga_{11}}  \\
       \partial_{\psi_{00}} & -\partial_{\psi_{10}}+(f_1-f_2)\partial_{\ga_{10}} & \partial_{\psi_{01}} &-\partial_{\psi_{11}}+(f_1-f_2)\partial_{\ga_{11}}
     \end{array}\nn\eea
Identifying $\partial_{\phi_{00}}=\partial_{\phi_{10}}=[0,0,1,0]$, $\partial_{\phi_{01}}=\partial_{\phi_{11}}=[0,0,0,1]$, $\partial_{\psi_{00}}=\partial_{\psi_{01}}=[1,0,0,0]$, $\partial_{\psi_{10}}=\partial_{\psi_{11}}=[0,-1,0,0]$, and taking into account (\ref{alg_relation}) and some normalisation, this table is identical to table (\ref{equi_action}).

Finally, the Reeb vector in the original coordinates is
\bea \reeb=\frac{\partial}{\partial\psi}-\frac16\frac{\partial}{\partial\ga}~.\nn\eea
Rewriting this in the new coordinates gives for example in patches $U_{00},~U_{10}$
\bea
 \reeb&=&\frac{\partial}{\partial\psi_{00}}+(f_2-\frac16)\frac{\partial}{\partial\ga_{00}}\nn\\
&=&-\frac{\partial}{\partial\psi_{10}}+(f_1-\frac16)\frac{\partial}{\partial\ga_{10}}~.\nn\eea
Close to $y=y_{1,2}$, the first terms both vanish, and the second terms $f_{2,1}-1/6=-y_{2,1}/6$ change sign. This fact plays a crucial role in method II of the index calculation.


\section{Calculation of the Index Method II}\label{sec_method_II}

This approach is less demanding on one's knowledge of the global geometry on the spcae in question but is certainly more long winded. The material we present here is in lecture 8 of \cite{Ellip_Ope_Cpct_Grp}. As was explained earlier, a transversally elliptic operator $D$ induces an element $[\gs(D)]\in K_G(T_G^*X)$, and the ellipticity ensures that the symbol $\gs(D)$ is an isomorphism except along the zero section of $T_G^*X$. It is convenient to pick a $G$-invariant metric and identify $T_G^*X$ with $T_GX$, consisting of the tangent vectors orthogonal to the group action. One can deform the symbol of the operator slightly along a vector field, and make the symbol into an isomorphism even on the zero section of $T_GX$, provided one stays away from the zero of the vector field in question. It will be natural to use a vector field generated by the group action for this job.

Now let us take $G=U(1)^n$, and define a filtration of $X$
\bea
X=X_{0}\supset X_1\supset\cdots \supset X_{n+1}=\emptyset~,\nn
\eea
where $X_i=\{x\in X:~\dim G_x\geq i\}$ with $G_x$ being the stability group at $x$. The key to the localisation of the index calculation is the following exact sequence
\bea
0\to K_G(T_G(X-X_i))\to K_G(T_G(X-X_{i+1}))\stackrel{\gt_i}{\leftrightharpoons} K_G(T_GX\big|_{X_i-X_{i+1}})\to0~,\nn\eea
The sequence above is split exact, and we will describe the splitting map $\gt_i$ shortly. From the split exactness, the middle term is just the direct sum of the two terms on the end
\bea K_G(T_G(X-X_{i+1}))\simeq K_G(T_G(X-X_i))\oplus\gt_i K_G(T_GX\big|_{X_i-X_{i+1}}).\nn\eea
One can continue with the same game with $K_G(T_G(X-X_i))$ and bootstrap oneself all the way from $X-X_1$ to $X-X_{n+1}=X$.

One can further determine $K_G(T_GX\big|_{X_i-X_{i+1}})$ by relating it to
$K_G(T_G(X_i-X_{i+1}))$ through the Thom isomorphism, since $T_GX\big|_{X_i-X_{i+1}}$ is a complex vector bundle over $T_G(X_i-X_{i+1})$. To summarise, we have that $K_G(T_GX)$ is a direct sum
\bea
K_G(T_GX)=\oplus_i \phi_iK_G(T_G(X_i-X_{i+1}))~,\label{K_direct_sum}
\eea
where $\phi_i$ is the composition of the Thom isomorphism and $\gt_i$. Hence given any symbol $[D]$ for which we want to compute the index, we can decompose $[D]\in K_G(T_GX)$ as the sum of classes $[D]=\phi_iK_G(T_G(X_i-X_{i+1}))$, in the hope that the index homomorphism of the summands can be determined by other means. In our situation, we will only need the cases when $X_i-X_{i+1}$ are either circles or points, and their $K$-group is
\bea
 K_G(pt)=R(G)~,~~~~K_{U(1)}(T_{U(1)}S^1)=K_{U(1)}(S^1)=R(U(1))/(\gl)~,~~~\gl=1-t^{-1}~,\label{index_hom_S1}
 \eea
where we use a Laurent polynomial $f(t)$ to denote both a function on $U(1)$ and direct sum of representations, for example $f(t)=t^{-1}+2t^2$ means a direct sum of 3 representations of $U(1)$, one with charge $-1$ and two with charge $2$. The index homomorphism is given by assigning a representation to its character, in particular for the zero symbol $[0]\in K_{U(1)}(T_{U(1)}S^1)$, its index is
\bea
\ind[0]=\gd(1-t)~,\label{ind_S^1}
\eea
which is annihilated by $(1-t^{-1})$, as it should from (\ref{index_hom_S1}).

Amongst the two parts making up $\phi$, the Thom isomorphism is well-understood, what is tricky is $\gt_i$. For clarity, we take a simpler situation, which is actually sufficient for all our calculations later
\bea
0\to K_G(T_G(\BB{C}-\{0\}))\to K_G(T_G\BB{C})\stackrel{\gt_{\pm}}{\leftrightharpoons} K_G(\BB{C})\to 0~,\nn
\eea
where $G=U(1)$ acting on $\BB{C}$ in the standard way\footnote{In this case $X=\BB{C}$, $X_1=\{0\}$, and $K_G(\{0\})\to K_G(\BB{C})=K_G(T_G\BB{C}\big|_{\{0\}})$ is the Thom isomorphism.\label{footnote_thom}}.
Take for example $1\in K_G(\{0\})$ and the Thom isomorphism multiplies to it the class $\gs=[\bar\partial]\in K_G(\BB{C})$.
One would like to 'insert' $\gs$ into $K_G(T_G\BB{C})$, in such a way that when restricted to $\BB{C}$ one gets $\gs$ back. One can certainly extend $\gs$ to $K_G(T_GU)$ where $U$ is a small neighbourhood of $\{0\}$ (using the retraction $U\to \{0\})$ for example). But this alone will not do, since in order to insert $\gs$ into $T_G\BB{C}$, one needs $\gs$ to have support only in a compact subset of $T_GU$ so as to make the insertion 'local', but $\gs$ is never an isomorphism on the zero section of $T_GU$, then as $U$ is open, the support of $\gs$ is not compact. This is where one needs to use the vector field generated by $G$ to deform $\gs$ so that $\gs$ is an isomorphism outside a compact subset of $T_G{U}$. Depending on in which direction one deforms $\gs$, one gets two classes $[\bar\partial^{\pm}]$, and if one applies the index homomorphism to these two classes, one gets
\bea \textrm{ind}_{U(1)}([\bar\partial^{\pm}])=\Big[\frac{1}{1-s^{-1}}\Big]^{\pm}~,\label{bar_partial_pm}\eea
where $s$ is the coordinate of $G=U(1)$. The superscript $\pm$ corresponds to two regularisations that send the rational function $1/(1-s^{-1})$ into distributions such that
\bea (1-s^{-1})\Big[\frac{1}{1-s^{-1}}\Big]^{\pm}=1~.\nn\eea
The explicit expression for $\textrm{ind}([\bar\partial^{\pm}])$ is
\bea &&P(s)=\Big[\frac{1}{1-s^{-1}}\Big]^+=-s-s^2-\cdots~,\nn\\
&&N(s)=\Big[\frac{1}{1-s^{-1}}\Big]^-=1+s^{-1}+s^{-2}+\cdots~,\nn\\
&&\gd(1-s)=N(s)-P(s)~,\nn\eea
it is easy to verify $(1-s^{-1})N(s)=(1-s^{-1})P(s)=1$ and hence $(1-s^{-1})\gd(1-s)=0$ as delta function does.

As an example (and a rather pedestrian one, but it will help us establish some conventions), let us compute the index of $[\bar\partial]\in K_G(T_G^*{\cal O}(p))$ where ${\cal O}(p)$ is the total space of the $U(1)$ bundle over $S^2$ of degree $p\geq0$, which can be presented as
\bea [z_1,z_2,e^{i\gt}]/\sim~,~~~~[z_1,z_2,e^{i\gt}]\sim[z_1e^{i\ga},e^{i\ga}z_2,e^{ip\ga+i\gt}]~.\nn\eea
The group $G$ is $U(1)^2$ with the action
\bea
s:~[sz_1,z_2,e^{i\gt}]~,~~~~u:~[z_1,z_2,ue^{i\gt}]~.\label{U1_action_north}
\eea
Note the second $U(1)$ is free. The above is valid close to the north pole $z_1=0$, while at the south pole,
the action of the two $U(1)$'s becomes
\bea
s:~[z_1,s^{-1}z_2,s^{-p}e^{i\gt}]~,~~~~~u:~[z_1,z_2,ue^{i\gt}]~.\label{U1_action_south}
\eea

It is not hard to decompose $[\bar\partial]$ into the direct sum (\ref{K_direct_sum}). At $z_1=0$ or $z_2=0$ one of the $U(1)$ degenerates, by applying the restriction map to, say, $z_1=0$, then $K_G(T_G{\cal O}(p))\to K_G(T_G{\cal O}(p)|_{z_1=0})=K_G(\BB{C}\times S^1)$. In this process $[\bar\partial]$ restricts to the 0 symbol on the $S^1$ factor, so we just get the contribution $[\bar\partial^{\pm}]\cdotp\gd(1-u)$ from the fixed point $z_1=0$, depending on our choice of the splitting map $\gt^{\pm}$. One gets a similar contribution $\gt^{\pm}K_G(\BB{C}\times S^1)$ from $z_2=0$, where the $U(1)^2$ actions on the two factors should be read off from (\ref{U1_action_south}).

How do we decide consistently which splitting map $\gt^{\pm}$ to use?
We will choose a global vector field generated by a $U(1)$-action, e.g. the first $U(1)$ in (\ref{U1_action_north}), and trivialise the symbol $[\bar\partial]$ globally except at the zeros of the vector field, then by scrutinising the symbol $[\bar\partial]$ close to the zeros, we can find out whether it should be $[\bar\partial^+]$ or $[\bar\partial^-]$.

The $U(1)$ we have chosen is just the standard rotation of a sphere. Denote the inhomogeneous coordinate of $S^2$ as $z=z_1/z_2$ and the vector field of this rotation is $v=i(z\partial-\bar z\bar\partial)$. Recall that the symbol $\gs(\bar\partial)$ sends a tangent vector $X$ to a bundle morphism $\gs(\bar\partial)(X):~\Go^{0,0}(S^2,{\cal O}(p))\to\Go^{0,1}(S^2,{\cal O}(p))$, one can deform the symbol into
\bea
 \gs^{\pm}(\bar\partial)(X)=\gs(\bar\partial)(X\pm v)~,\nn
 \eea
which will remain a bundle isomorphism even when $X=0$, thus trivialising the symbol $\gs(\bar\partial)$ except at the two poles. This deformation correspond to the symbol $[\bar\partial^{\pm}]$ introduced earlier. We choose the $+$ option (choosing $-$ makes no difference).

Now we can assemble the contributions from the two poles (using also (\ref{bar_partial_pm}))
%
\bea
\textrm{north pole}:~~ \Big[\frac{1}{1-s^{-1}}\Big]^+\gd(1-u)=-(s+s^2+\cdots)\gd(1-u)~,\nn\\
 \textrm{south pole}:~~ \Big[\frac1{1-s}\Big]^+\gd(1-us^{-p})=(1+s+s^2+\cdots)\gd(1-us^{-p})~.\nn\eea
Combining the the two terms, and we collect the coefficient proportional to $u^n$
\bea \ind([\bar\partial])(u^n)=-(s+s^2+\cdots)+(1+s+s^2+\cdots)s^{-pn}=\Big\{
                                                                              \begin{array}{cc}
                                                                                n\geq0 & 1+\cdots+s^{-pn} \\
                                                                                n<0 & -s-\cdots -s^{-pn-1} \\
                                                                              \end{array}~,\nn\eea
in particular, evaluating at $s=1$, the coefficient of $u^n$ is the index
\bea
\dim H^0(S^2,{\cal O}(pn))-\dim H^1(S^2,{\cal O}(pn))~.\nn
\eea
Notice that in general one does not have the luxury of evaluating an index at certain value, since the indices will be a distribution not a function.

Now we are ready to take on our problem of computing the index on $Y^{p,q}$. We take $G=U(1)^3$ generated by $e_1,e_3$ of table (\ref{equi_action}) and $\ga$ which is free, and denote by $s,t$ and $u$ the coordinates of the three $U(1)$'s.

The operator $D$ we want to compute is the $\bar\partial$-operator whose complex structure is the one determined by $J$, which as we saw in sec.\ref{sec_U(1)_base} agrees with one on the base $S^2$ of the fibration $S^2\rtimes S^2$. While over the fibre $S^2$, it is homotopic to $\bar\partial$ close to the north pole, but to $-\partial$ close to the south pole.
We will again choose $e_1$ and $e_3$ to trivialise the symbol $\gs(D)$ everywhere except the four poles, and from the previous discussion of the complex structure, the trivialisation procedure is identical to the previous example when restricted to the base $S^2$. On the fibre sphere, the situation is similar to the case in lemma 6.4 in \cite{Ellip_Ope_Cpct_Grp}, and one can trivialise $D$ by trivialising $\bar\partial$ with the positive regularisation close to the north pole and $-\partial$ with the negative regularisation close to the south pole. This choice of regularisation will then mesh together at the equator.

To summarise, we have the following contribution (where the notation of the four patches $U_{ij}$ were defined in subsection \ref{sec_U(1)_base})
\bea \textrm{pole in } U_{00}:&&\Big[\frac{1}{1-s^{-1}}\Big]^+\Big[\frac{1}{1-t^{-1}}\Big]^+\gd(1-u)~,\nn\\
 \textrm{pole in } U_{01}:&&\Big[\frac{1}{1-v^{-1}}\Big]^+\Big|_{v=t^2s}\Big[\frac{1}{1-t}\Big]^+\delta(1-ut^{q-p})~,\nn\\
\textrm{pole in } U_{10}:&&\Big[\frac{1}{1-s^{-1}}\Big]^-\Big[\frac{1}{1-t^{-1}}\Big]^+\delta(1-us^p)~,\nn\\
\textrm{pole in } U_{11}:&&\Big[\frac{1}{1-v^{-1}}\Big]^-\Big|_{v=t^2s}\Big[\frac{1}{1-t}\Big]^+\delta(1-us^pt^{p+q})~,\nn\eea
where the change of variable can be read off from the table (\ref{equi_action}).

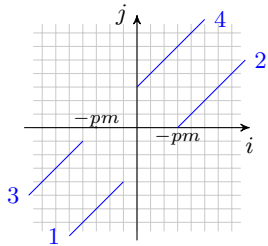
\begin{figure}[h]
\begin{center}
\begin{tikzpicture}[scale=.6]
\draw [step=0.3,thin,gray!40] (-2.3,-2.3) grid (2.3,2.3);

\draw [->] (-2.5,0) -- (2.5,0) node [below] {\small$i$};
\draw [->] (0,-2.5) -- (0,2.5) node [left] {\small$j$};

\draw [-,blue] (0,.9) -- (1.5,2.4) node[right] {\small$4$};
\draw [-,blue] (0.9,0) -- (2.4,1.5) node[right] {\small$2$};

\draw [-,blue] (-0.3,-1.2) -- (-1.5,-2.4) node[left] {\small$1$};
\draw [-,blue] (-1.2,-0.3) -- (-2.4,-1.5) node[left] {\small$3$};

\node at (.9,-.2) {\scriptsize$-pm$};
\node at (-.9,.2) {\scriptsize$-pm$};
\node at (-.9,0) {\scriptsize$\cdot$};
\end{tikzpicture}\caption{The two lower lines have $m<0$ and the two upper $m\geq0$. And each lattice point on the lines in the first quadrant has multiplicity $1+2i+m(p-q)$, while those in the third quadrant $-2i+m(q-p)-1$.}\label{lattice_comparison}
\end{center}
\end{figure}

Now combine the $U_{00}$ and $U_{10}$ contribution, and single out the term of power $u^m$, then the sum over $s$ looks like
\bea m<0:~~\Big(-\sum_1^{\infty}s^k+\sum^{pm}_{-\infty}s^k\Big)\Big[\frac{1}{1-t^{-1}}\Big]^+,~~~~~m\geq0:~~\Big(-\sum^{\infty}_{pm+1}s^k+\sum^0_{-\infty}s^k\Big)\Big[\frac{1}{1-t^{-1}}\Big]^+.\nn\eea
The combination of $U_{10}$ and $U_{11}$ contribution can be obtained by replacing $s\to t^2s$, $t\to t^{-1}$ as well as including an overall factor $t^{m(q-p)}$
\bea m<0&& t^{m(q-p)}\Big(-\sum_1^{\infty}(t^2s)^k+\sum^{pm}_{-\infty}(t^2s)^k\Big)\Big[\frac{1}{1-t}\Big]^+\nn\\
m\geq0&& t^{m(q-p)}\Big(-\sum^{\infty}_{pm+1}(t^2s)^k+\sum^0_{-\infty}(t^2s)^k\Big)\Big[\frac{1}{1-t}\Big]^+.\nn\eea
Once all four contributions are combined, one observes that one can send $t=1$ safely, this is so because $D$ restricts to an elliptic operator on the base sphere. We will keep $t$ and expand out $[1/(1-t^{\pm 1})]^+$, there will be some partial cancellations, which correspond to the missing segments between line 1,2 and 3,4 in fig.\ref{lattice_comparison}.
\bea m<0&& \Big(\sum_1^{\infty}s^k\sum_{j=1}^{2k+m(q-p)-1}t^j+\sum^{pm}_{-\infty}s^k\sum_{2k+m(q-p)}^0t^j\Big)\nn\\
m\geq0&&\Big(\sum^{\infty}_{pm+1}s^k\sum_{j=1}^{2k+m(q-p)-1}t^j+\sum^0_{-\infty}s^k\sum_{2k+m(q-p)}^0t^j\Big)~.\nn\eea
To facilitate comparison with the calculation in subsection \ref{sec_method_I}, we rename $k=-i$, $j=-k$ and our index becomes
\bea m<0&& \Big(\sum^{-1}_{i=-\infty}s^{-i}\sum^{-1}_{k=2i-m(q-p)+1}t^{-k}+\sum_{i=-pm}^{\infty}s^{-i}\sum^{2i-m(q-p)}_{k=0}t^{-k}\Big)\nn\\
m\geq0&&\Big(\sum_{i=-\infty}^{-pm-1}s^{-i}\sum^{-1}_{k=2i-m(q-p)+1}t^{-k}+\sum_{i=0}^{\infty}s^{-i}\sum^{2i-m(q-p)}_{k=0}t^{-k}\Big)~.\nn\eea
The four terms correspond each to the four lines in the fig.\ref{lattice_comparison}, and the agreement with (\ref{index_M_I}) is crystal clear. As we have already remarked in (\ref{mode_alpha}) that the mode $m$ in (\ref{only_if}) is the mode of the free $U(1)$ denoted as $\ga$, and here by doing the calculation differently, we are merely changing the order of summation (but in a manner that is allowed for an infinite sum).


\providecommand{\href}[2]{#2}\begingroup\raggedright\endgroup

\end{document}